\documentclass[12pt,preprint]{aastex}

\usepackage{epsfig}
\usepackage{amsmath}
\usepackage{amssymb}
\usepackage{amsbsy}
\usepackage{placeins}
\usepackage{color}
\usepackage{lscape}

\newcommand{\D}{\mbox{d}}

\newcommand{\iles}{{\sc iles}}

\newcommand{\Fr}{\mathrm{Fr}}

\usepackage{ulem}
\newcommand{\gtaprx}{\lower .1ex\hbox{\rlap{\raise .6ex\hbox{\hskip .3ex
        {\ifmmode{\scriptscriptstyle >}\else
                {$\scriptscriptstyle >$}\fi}}}
        \kern -.4ex{\ifmmode{\scriptscriptstyle \sim}\else
                {$\scriptscriptstyle\sim$}\fi}}}
\newcommand{\ltaprx}{\lower .1ex\hbox{\rlap{\raise .6ex\hbox{\hskip .3ex
        {\ifmmode{\scriptscriptstyle <}\else
                {$\scriptscriptstyle <$}\fi}}}
        \kern -.4ex{\ifmmode{\scriptscriptstyle \sim}\else
                {$\scriptscriptstyle\sim$}\fi}}}

\usepackage{epsf,color,dcolumn}

\newcolumntype{d}{D{.}{.}{-1}}

\epsfverbosetrue


\begin{document}

\title{Burning Thermals in Type Ia Supernovae}

\author{A.~J.~Aspden\altaffilmark{1}, J.~B.~Bell\altaffilmark{1}, S.~Dong\altaffilmark{2}, and S.~E.~Woosley\altaffilmark{2}}

\altaffiltext{1}{Lawrence Berkeley National Laboratory, 1 Cyclotron
Road, MS 50A-1148, Berkeley, CA 94720}
\altaffiltext{2}{Department of Astronomy and Astrophysics, University
of California at Santa Cruz, Santa Cruz, CA 95064}

\begin{abstract}
We develop a one-dimensional theoretical model for thermals burning in Type Ia 
supernovae based on the entrainment assumption of Morton, Taylor and Turner.  
Extensions of the standard model are required to account for the burning and for the 
expansion of the thermal due to changes in the background stratification found in the 
full star.  The model is compared with high-resolution three-dimensional numerical 
simulations, both in a uniform environment, and in a full-star setting.
The simulations in a uniform environment present compelling agreement with 
the predicted power-laws and provide model constants for the full-star model,
which then provides excellent agreement with the full-star simulation.  The importance
of the different components in the model are compared, and are all shown to be
relevant.  
An examination of the effect of initial conditions was then conducted using the 
one-dimensional model, which would have been infeasible in three dimensions.  
More mass was burned when the ignition kernel was larger and
closer to the center of the star.  The turbulent flame speed was found to be important 
during the early-time evolution of the thermal, but played a diminished role at later 
times when the evolution is dominated by the large-scale hydrodynamics responsible for 
entrainment.
However, a higher flame speed effectively gave a larger initial ignition
kernel and so resulted in more mass burned.  This suggests that future studies should 
focus on the early-time behavior of these thermals (in particular, the transition to 
turbulence), and that the choice of turbulent flame speed does not play a significant
role in the dynamics once the thermal has become established.
\end{abstract}

\section{INTRODUCTION}
\label{sec:intro}

Type Ia supernovae are among the brightest explosions in the modern
universe and are a useful tool for cosmological distance
determination. It is generally agreed that they result from the
thermonuclear explosion of a white dwarf accreting matter from a
companion star in a binary system \citep{Hoy60}. The more specific
nature of the progenitor though, be it a Chandrasekhar-mass white
dwarf, sub-Chandrasekhar mass dwarf, or even two merging white
dwarfs, is debated \citep[e.g.][]{hillebrandtniemeyer2000,Yoo07,Sim10}, though most
recent work has focused on the ``single'' Chandrasekhar-mass
progenitor scenario.

Whatever the progenitor may be, the fact that a thermonuclear
explosion is involved means that a realistic model requires an
understanding of both the ignition and propagation of the burning.
Barring a sudden detonation which, at least for the Chandrasekhar mass
model, is ruled out by observations, the runaway proceeds through
several stages. First compression raises the temperature to a point
where local nuclear energy generation exceeds losses, e.g., by
neutrino emission. Given the degeneracy of the electrons, the
temperature continues to rise with the excess energy being carried
away by convection. This phase may last centuries. Finally, subsonic
convection is unable to carry all the energy and the temperature
gradient steepens and becomes grossly superadiabatic. Again absent a
detonation, the gradient continues to steepen until a small region
runs away to very high temperature creating a sharp interface between
fuel and ash called a ``flame''. The subsequent rapid propagation of
this flame causes the supernova.

In this paper, we focus on the Chandrasekhar mass model, though there
are probably generalizations to the other models. Recent studies
\citep[e.g.][]{Zin09,Zin11} suggest that this sort of supernova will
not usually ignite at its center. Instead, the initial burning is
offset to one side by 30 - 90 km. Though small in comparison with the
white dwarf size, about 1700 km, this displacement is sufficient for
the supernova to burn only on one side \citep{Nie96,Zin07,Roepke07,Jor08}. Because of
the extreme temperature sensitivity of the nuclear energy generation,
the burning probably begins as one or a few isolated hot spots, which
we will represent here as a small sphere of ash. The surface of this
sphere is deformed, perhaps grossly so, by the turbulence generated
during the pre-explosive convection. Because of the burning and high
temperature, the density of the bubble, under isobaric conditions, is
smaller than its surroundings. It therefore floats and might be
described as a ``bubble''. As the bubble rises, instabilities develop
on its perimeter while entrainment, burning, and expansion increase
its mass and volume.  Three-dimensional simulations of the bubble's
behavior are presented for physical conditions appropriate to the
supernova, and an analytic model is developed to explain its late
time evolution.  The answers obtained are very important to
understanding the outcome of Type Ia supernova explosion
models. Indeed, much of the diversity observed in Type Ia
supernovae, both on the computer \citep{Roepke07,Jor08} and in nature
\citep{Kasen09}, probably results from different geometries of the
ignition region and evolutions of the first plumes of burning as they
rise to the surface of the star.  We illustrate and validate our
analytic model by comparing to a three-dimensional full-star model of
a supernova \citep{Dong11}.

Much use is made here of the rich literature in fluid mechanics
concerning similar phenomena in the terrestrial context.
In that literature, a localized blob of buoyant fluid is
known as a thermal, which is considered a special case of a buoyant
vortex ring; the difference being that the latter has initial
momentum in addition to its buoyancy.  \cite{Scorer57} investigated
thermals experimentally, and proposed relations based on dimensional
analysis of the form 
\begin{equation}
z=nr\quad\mathrm{and}\quad w=C(g\bar{B}r)^{1/2},
\label{Eq:Scorer}
\end{equation}
where (using Scorer's notation) $z$ and $r$ are the height and radius
of the thermal, $\bar{B}$ the mean buoyancy, $w$ the vertical
velocity, $g$ acceleration due to gravity, and $n$ and $C$ are
constants, where the latter is of the form of a Froude number.  

\citet*{Morton56} introduced the concept of an entrainment assumption.
This approach considers an idealized thermal at time $t$, represented 
by a sphere of radius $b(t)$ at a height $z(t)$, traveling vertically 
with speed $u(t)=\D z/\D t$.  The entrainment assumption states that 
the entrainment at the edge of a thermal is proportional to some 
characteristic velocity within the thermal, which is usually taken to be $u$.
Under the Boussinesq approximation, conservation equations for volume 
(or equivalently mass), momentum and buoyancy can be written as 
\begin{eqnarray}
\frac{\D}{\D t}\left(\frac{4}{3}\pi b^3\right) &=& 
4 \pi b^2 \alpha u,\label{Eq:MttVol}\\
\frac{\D}{\D t}\left(\frac{4}{3}\pi b^3 (k_v\rho_e+\rho_i) u\right) &=& 
\frac{4}{3} \pi b^3 \beta \left(\rho_e-\rho_i\right) g,\label{MttMom}\\
\frac{\D}{\D t}\left(\frac{4}{3}\pi b^3  g \frac{\rho_e-\rho_i}{\rho_0}\right) &=& 
-\frac{4}{3} \pi b^3 u N^2,\label{Eq:MttBuoy}
\end{eqnarray}
where $\alpha$ is the constant entrainment coefficient, $\rho_i$,
$\rho_e$ and $\rho_0\equiv\rho_e(0)$ are the densities of the
interior of the thermal, the ambient fluid and a reference density,
respectively, and $N^2=-(g/\rho_0)(\D\rho_e/\D z)$ is the
Brunt-V\"ais\"al\"a buoyancy frequency describing the ambient
stratification.  
The first equation states that the rate of change of the total 
volume of the thermal is equal to the volume of the entrained fluid,
i.e.\ the entrainment rate $\alpha u$ times the surface area 
of the thermal.  This is equivalent to rate of change of mass
under the Boussinesq approximation.  The second equation is
just Newton's second law, and states that the change of 
momentum is equal to the total buoyancy force acting on 
the thermal.  The `virtual mass coefficient' $k_v$ has been 
included here to account for the momentum of the ambient fluid
surrounding the thermal, which takes the theoretical value of one half for 
a rigid sphere, and we have also included an empirical constant $\beta$.  
The last equation describes the rate of change of buoyancy due to the 
density of the entrained fluid and background stratification.
It follows immediately from equation (\ref{Eq:MttVol}) that 
$b=\alpha z$, for a suitably defined origin, as in equation
(\ref{Eq:Scorer}).  This means that the thermal prescribes a cone as it
rises, regardless of the ambient conditions.  For an unstratified
ambient, $\rho_e\equiv\rho_0$ and $N^2\equiv0$, the velocity from
equation (\ref{Eq:Scorer}) can be easily derived.

Motivated by latent heat release in atmospheric clouds,
\cite{Turner63} considered thermals where the buoyancy was enhanced 
due to chemical reaction.  The experiments and analysis were
restricted to two kinds of motion, either constant acceleration or
constant velocity.  We will see that the former case is relevant to
burning thermals in SN Ia.  A review of these early works can be found 
in \cite{Turner69}. 

In this paper, we consider applying this simple entrainment
approach to an off-center ignition in SN Ia.  Several effects 
have to be taken into consideration.  The Boussinesq
approximation is no longer appropriate due to the large
variation in density.  The effects of burning have to be 
considered, in particular changes in density and volume 
(and therefore buoyancy).  Finally, the background stratification
in a full star cannot be represented by a constant buoyancy
frequency $N$, or even constant acceleration due to gravity $g$.  
We first consider the effects of burning in a uniform ambient 
to establish the applicability of the entrainment assumption 
to burning thermals, and establish values for the constants
$\alpha$ and $\beta$.  A theoretical description is given and 
compared with new high-resolution numerical simulations.  
We then consider a burning thermal in a full-star setting.
A theoretical description is given that extends the model to
a non-uniform background, which is then compared with the high-resolution 
numerical simulations conducted by \cite{Dong11}.  Each component
of the one-dimensional model is examined and shown to be important
to capture the behavior of the burning thermal in a full-star
environment.  The effect of initial conditions are then examined with
the one-dimensional model by varying the initial radius, height, and 
flame speed of the thermal.  More mass was burned when the ignition 
kernel was larger and closer to the center of the star.  The turbulent 
flame speed was found to be important during the early-time evolution 
of the thermal, but played a diminished role at later times when the 
evolution is dominated by the large-scale hydrodynamics responsible 
for entrainment.  However, a higher flame speed gave the 
effect of a larger initial ignition kernel and so resulted in more 
mass being burned.

\section{THERMAL BURNING IN A UNIFORM AMBIENT}

\subsection{Analysis}
\label{Sec:Analysis}

To modify the analysis to account for the continual buoyancy
increase due to heat release during burning, we assume that 
entrained fluid burns sufficiently quickly
that the density inside the
thermal can be treated as roughly constant in both time and space.
Specifically, we can use the `top hat' approach discussed in
\cite{Morton56}, i.e.\ any point in space can be simply inside or
outside the thermal.  For uniform ambient, the outside density is
a constant equal to the fuel density and burns to a constant ash
density (instantaneously) upon entrainment.  This means that there 
can only be two discrete densities, fuel $\rho_F$ and ash $\rho_A$.
This simplifies the equations as the Boussinesq approximation is
no longer required (although the equations for volume and mass 
remain equivalent here because the densities are constant) and 
the buoyancy equation is redundant.  Under these assumptions, the
conservation equations (\ref{Eq:MttVol})-(\ref{Eq:MttBuoy}) reduce 
to two equations for volume and momentum
\begin{eqnarray}
\frac{\D}{\D t}\left(\frac{4}{3}\pi b^3\right) &=& 
4 \pi b^2 \sigma \alpha u,\label{eq:vol}\\
\frac{\D}{\D t}\left(\frac{4}{3}\pi b^3 (k_v\rho_F+\rho_A) u\right) &=& 
\frac{4}{3} \pi b^3 \beta \left(\rho_F-\rho_A\right) g\label{eq:mom},
\end{eqnarray}
where $\sigma=\rho_F/\rho_A$ is the ratio of fuel and ash densities,
and the acceleration due to gravity $g$ is assumed to be a constant.
Note the density ratio $\sigma$ in equation (\ref{eq:vol}) accounts for the 
expansion due to burning.  Equivalently, the $\rho_A$ can be included 
in the derivative on the left-hand size yielding a mass conservation 
equation.  Again, note that the Boussinesq 
approximation has not been made here.

As before, it follows immediately that $b=\sigma \alpha z$, for a 
suitably defined origin, and so the burning thermal also prescribes
a cone as it evolves.  Substituting into equation (\ref{eq:mom}) gives
\begin{equation}
\frac{\D^2 z}{\D t^2} + 
\frac{3}{z}\left(\frac{\D z}{\D t}\right)^2 
= g^\prime,
\label{eq:second}
\end{equation}
where $g^\prime=g\beta(\rho_F-\rho_A)/(k_v\rho_F+\rho_A)$ is a constant 
reduced gravity.
Equation (\ref{eq:second}) has the solution 
\begin{equation}
z=\frac{g^\prime}{14}t^2,
\quad\quad
u=\frac{g^\prime}{7}t,
\quad\quad\mathrm{and}\quad\quad
b=\sigma\alpha z=\frac{\sigma\alpha g^\prime}{14}t^2.
\label{eq:soln}
\end{equation}
Equation \ref{eq:soln} describes the self-similar evolution of
a burning thermal.
Note that this has the same form as the constant acceleration case
considered by \cite{Turner63}.  A Froude number can be defined as
\begin{equation}
\Fr^2=\frac{u^2}{g^\prime b},
\nonumber
\end{equation}
which in the self-similar regime takes the value $\Fr_\infty^2=2/(7\sigma\alpha)$.
Moreover, an evolution equation for the Froude number can be derived
\begin{equation}
\frac{\D}{\D t}\left(\Fr^2\right) = T\left(\Fr_\infty^2-\Fr^2\right),
\label{eq:SimpleFr}
\end{equation}
where $T=7\sigma\alpha u/b$.
This means that $\Fr=\Fr_\infty$ is a stable equilibrium for positive $u$, 
i.e.\ the Froude number will tend towards $\Fr_\infty$ at late times.

To assess the applicability of such an entrainment model, we now compare 
this theoretical treatment with numerical simulations of a burning thermal 
in a uniform environment where the flame physics are well-resolved.

\subsection{Numerical Approach}
\label{Sec:Numerics}

We use a low Mach number hydrodynamics code, adapted to
the study of thermonuclear flames, as described in \citet{SNeCodePaper}.
The advantage of this method is that sound waves are filtered out
analytically, so that the time step is set by the the bulk fluid velocity
and not the sound speed.  This is an enormous efficiency gain for low
speed flames.  The input physics used in the present simulations is
largely unchanged (we consider only carbon burning to magnesium), with 
the exception of the addition of Coulomb
screening, taken from the Kepler code \citep{weaver:1978}, to the
$^{12}$C($^{12}$C,$\gamma$)$^{24}$Mg reaction rate.  This yields a
small enhancement to the flame speed, and is included for
completeness.  The conductivities are those reported in
\citet{timmes_he_flames:2000}, and the equation of state is the
Helmholtz free-energy based general stellar EOS described in
\citet{timmes_swesty:2000}.  We note that we do not utilize the
Coulomb corrections to the electron gas in the general EOS, as these
are expected to be minor at the conditions considered. 

Adaptive mesh refinement is achieved through a block-structured approach that
uses a nested hierarchy of logically-rectangular grids with simultaneous 
refinement of the grids in both space and time, originally developed by
\citet{bergercolella,bell-3d}.  The integration algorithm on the
grid hierarchy is a recursive procedure in which coarse grids are
advanced in time, fine grids are advanced multiple steps to reach the
same time as the coarse grids and the data at different levels are
then synchronized.  During the regridding step, increasingly finer
grids are recursively embedded in coarse grids until the solution is
sufficiently resolved.  An error estimation procedure based on
user-specified criteria evaluates where additional refinement is
needed, and grid generation procedures dynamically create or remove
rectangular fine grid patches as resolution requirements change.
A coarse-grained parallelization strategy is used to distribute
grid patches to nodes where the nodes communicate using MPI.

The non-oscillatory finite-volume scheme employed here permits the use of
implicit large eddy simulation (\iles).  This technique captures the inviscid cascade
of kinetic energy through the inertial range, while the numerical error acts
in a way that emulates the dissipative physical effects on the dynamics at the grid
scale, without the expense of resolving the entire dissipation
subrange.  An overview of the technique can be
found in \cite{GrinsteinBook07}.  \cite{Aspden08b}
presented a detailed study of the technique using the present numerical
scheme, including a characterization that allowed for an effective viscosity
to be derived.  Thermal diffusion plays a significant role in the
flame dynamics, so it is explicitly included in the model, whereas
species diffusion is significantly smaller, so it is not included. 

The aim of this small-scale study is to establish the applicability
of the entrainment model for a burning thermal using simulations
where the flame physics are well-resolved.  We consider a 
carbon-burning thermal (50\% carbon, 50\% oxygen) in a cubic domain 
of side length 864\,cm with a free-slip 
base and outflow boundary conditions elsewhere.  The fuel density
was 1.5$\times10^7$\,g\,cm$^{-3}$, which burned to an ash density of 0.85$\times10^7$\,g\,cm$^{-3}$,
hence $\sigma=1.76$.  Gravity was held constant at $10^9$\,cm\,s$^{-2}$.
These conditions were chosen so that the flame was well-resolved at an 
accessible resolution, but the specific details are not expected to be 
important, as any self-similar evolution is independent of these factors.
The thermal was initialized 54\,cm above the bottom of the 
domain in the center, with a radius of approximately 14\,cm and a 
significant perturbation to the surface to break symmetry.
The density and temperature inside and outside the thermal were set 
equal to that of the ash and the fuel, respectively.  The velocity
was set to be zero throughout the domain.
Adaptive mesh refinement was used to focus resolution on the 
thermal, lowering computational expense.  Two simulations of burning thermals were
run, differing only in resolution.  The high resolution simulation,
which we will refer to as Run H, had a base grid of
$512^3$, with three levels of refinement, giving an
effective resolution of $4096^3$.  The theory predicts that 
the volume of the thermal grows with time to the sixth power, 
which means the refined region rapidly becomes large.  Once the 
thermal had grown such that the simulation was prohibitively
expensive, a level of refinement was removed, evolving the
thermal at an effective resolution of $2096^3$.  This step
was repeated such that the final effective resolution was 
$1024^3$.  We will refer to the three parts of the simulation
as stages 1 through 3.  The low resolution simulation, which
we will refer to as Run L, had an effective resolution of 
$2048^3$ from the beginning and was evolved through two stages 
in a similar manner to Run H.  A third simulation was run
under the same conditions at the same resolution as case H
with the burning turned off to provide a comparison with
conventional inert thermals.

\subsection{Results}
\label{Sec:Results}

\begin{figure}
\centering
\plotone{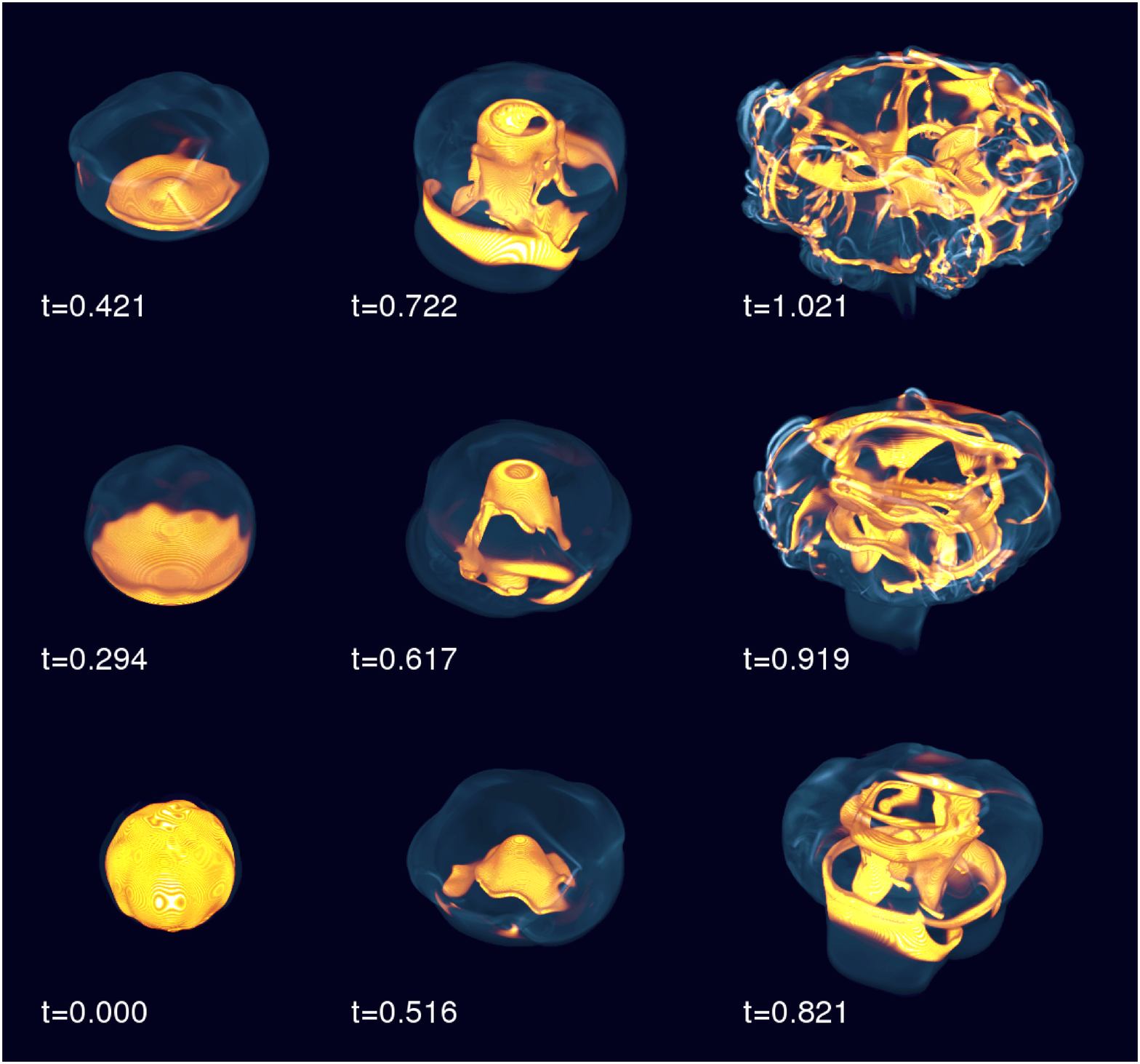}
\caption{Three-dimensional renderings of the early stages of evolution of a
burning thermal in a uniform environment.  The yellow denotes burning and blue denotes
vorticity.  All images are at the same scale, where the initial thermal radius was 14\,cm.
The times corresponding to each image is given in milliseconds (note for context the 
entire simulated time was 5.6\,ms, see figure \ref{Fig:late3d} for evolution after $t=1.021$\,ms).}
\label{Fig:early3d}
\end{figure}

The early stages of development are depicted in figure \ref{Fig:early3d}
by three-dimensional renderings of burning (yellow) and magnitude of
vorticity (blue) at nine time points between $t=0$ and $1.021$\,ms 
(for context the thermal reached to top of the domain after approximately
5.6\,ms).  All images are at the same scale and orientation; recall the 
initial radius is approximately 14\,cm in a domain size of 864\,cm.
In the initial conditions, the burning is uniformly distributed 
over the surface of the thermal, and there is zero vorticity as the
entire domain is stationary.  As the thermal begins to rise, burning at
the cap is extinguished by stretch induced by the large-scale flow 
around the thermal ($t=0.294$\,ms).  Thermodiffusive effects due to the
high Lewis numbers lead to intense burning on the underside of the 
thermal ($t=0.421$\,ms).  Where the surface is deformed in such a way 
that the center of curvature is within the fuel, thermal diffusion focuses
heat into the fuel, and so increases the burning rate.  The curvature is
maintained by the large-scale flow, and the intense burning region 
continues to burn upwards into the center of the thermal ($t=0.512$\,ms).
Eventually, this intense burning region is extinguished by the large-scale
flow internal to the thermal ($t=0.722$\,ms).  As the thermal toruses 
($t=0.821$\,ms), the burning also occurs in toroidal structures, before 
secondary shear instabilities initiate the transition to turbulence 
($t=0.919$--$1.021$\,ms).  

\begin{figure}
\centering
\plotone{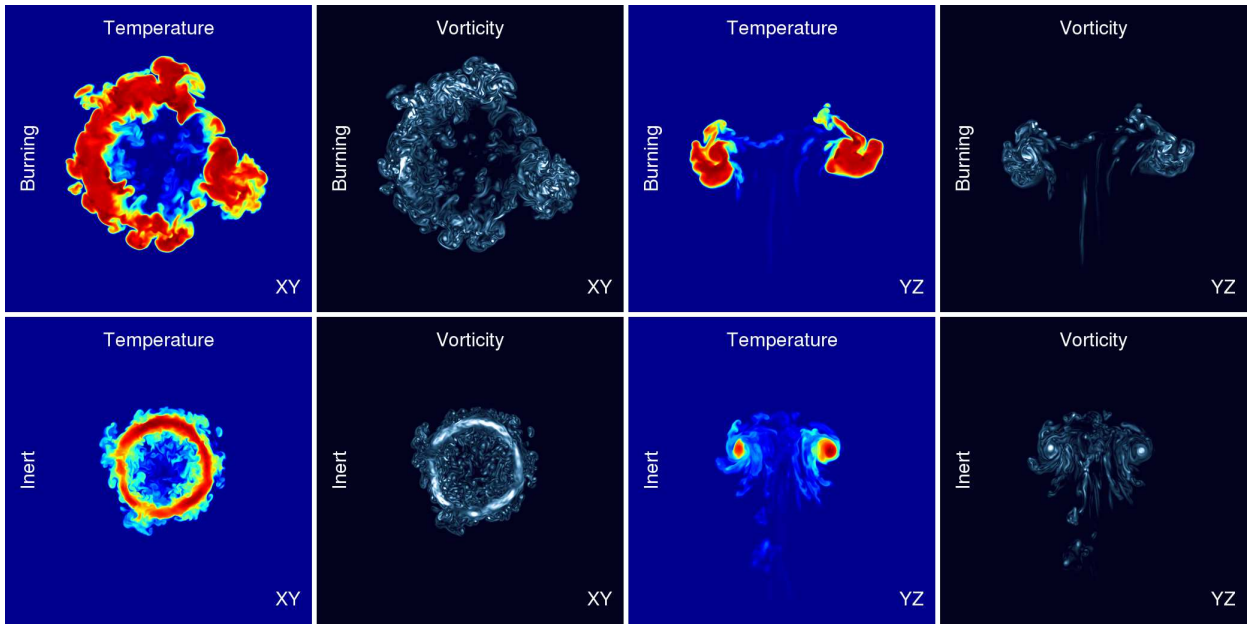}
\caption{Two-dimensional slices from the three-dimensional simulations
of a burning thermal (top) and an inert thermal (bottom) in a uniform ambient.
For temperature, red is hot, blue is cold.  For vorticity, white is high, black is low.
The time of the burning thermal is 2.042\,ms, and the inert thermal is taken at 
the same rise height.  All images are at the same scale.}
\label{Fig:earlySlices}
\end{figure}

Two-dimensional slices through the data shown in figure \ref{Fig:earlySlices}
compare the burning thermal (top row)
at $t=2.042$\,ms with the inert thermal (bottom row) at the same rise height.
The four left-hand panels are horizontal slices, and the four right-hand panels
are vertical slices. 
Both thermals have a toroidal structure, but the burning case has larger primary
and secondary radii, and is considerably more asymmetric.  The burning thermal
has an almost bimodal temperature, close to the ash temperature inside, and fuel
temperature outside.  The vorticity is confined to this hot region, and is not 
present in the center, the inert thermal has vorticity 
distributed across the entire structure.  The inert thermal has a strong and
persistent coherent toroidal vortex core, which is not present in the burning case.
Burning appears to suppress the transition to turbulence and self-similar evolution,
and the burning thermal expands sideways more rapidly than in the inert case.

\begin{figure}
\centering
\plotone{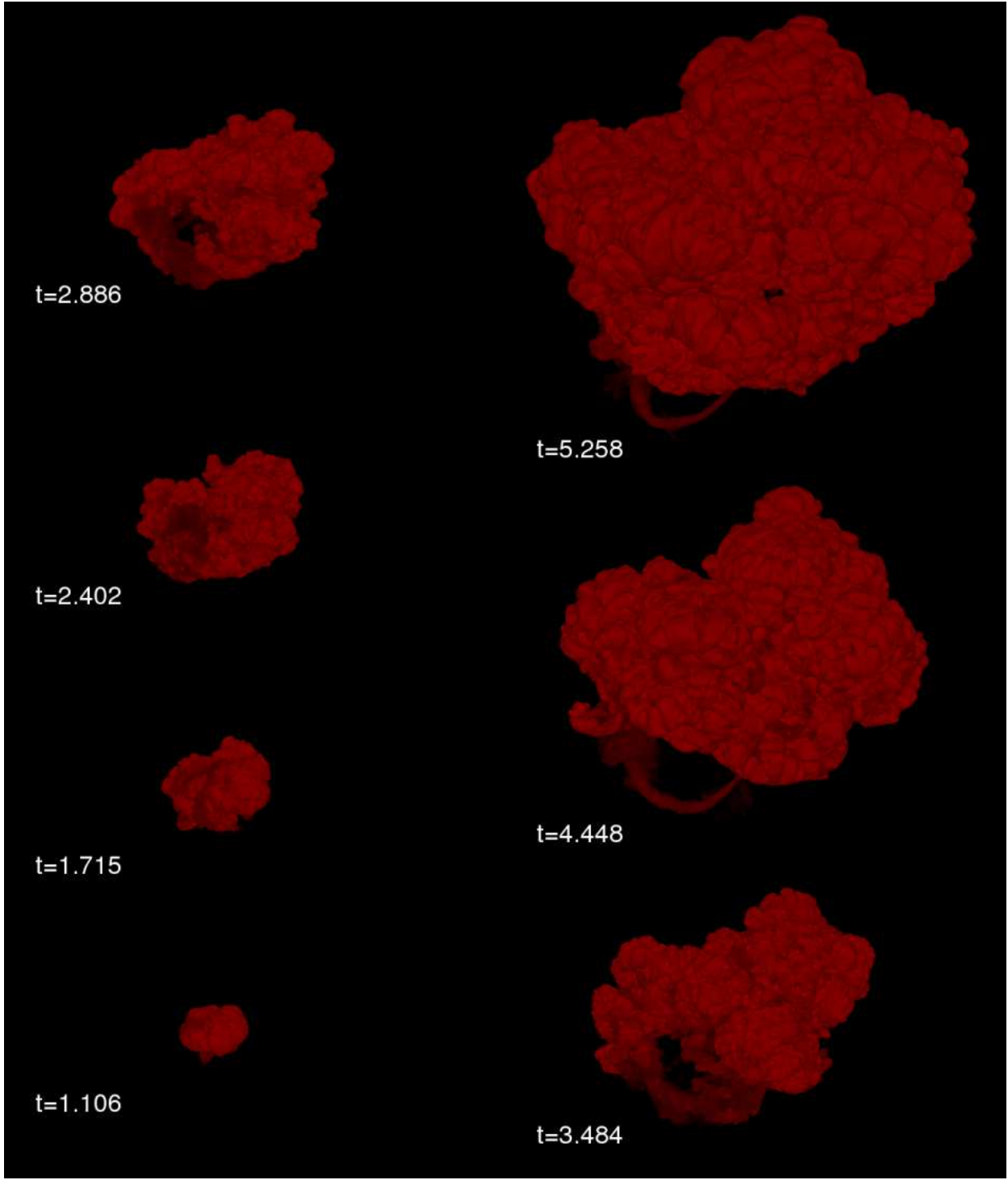}
\caption{Three-dimensional renderings of the later stages of evolution of 
the temperature of a burning thermal in a uniform environment.
All images are at the same scale and times are given in milliseconds.
Note the first time corresponds to the latest time shown in figure \ref{Fig:early3d},
and highlights the rapid growth of the thermal.  At times $t\ltaprx0.8$\,ms, the 
burning is mainly on the underside of the thermal, but at around $t\approx1$\,ms, 
the burning becomes distributed throughout the thermal.}
\label{Fig:late3d}
\end{figure}

Three-dimensional renderings of temperature in figure \ref{Fig:late3d} depict the
thermal evolution over the later times of the simulation.  All images are shown at the
same scale, and the first image is close in time (and therefore size) to the final 
image from figure \ref{Fig:early3d}, which, contrasted with the image at 
($t=5.258$\,ms) highlights the rapid growth of the thermal after the transition
to turbulence.  A local extinction event occurs around $t=2.886$\,ms (in the lower
left of the image as shown), which leaves behind a coherent vortex structure that 
persists for the remainder of the simulation.  However, the burning continues
throughout the rest of the thermal, which dominates the overall evolution.

\begin{figure}
\centering
\plotone{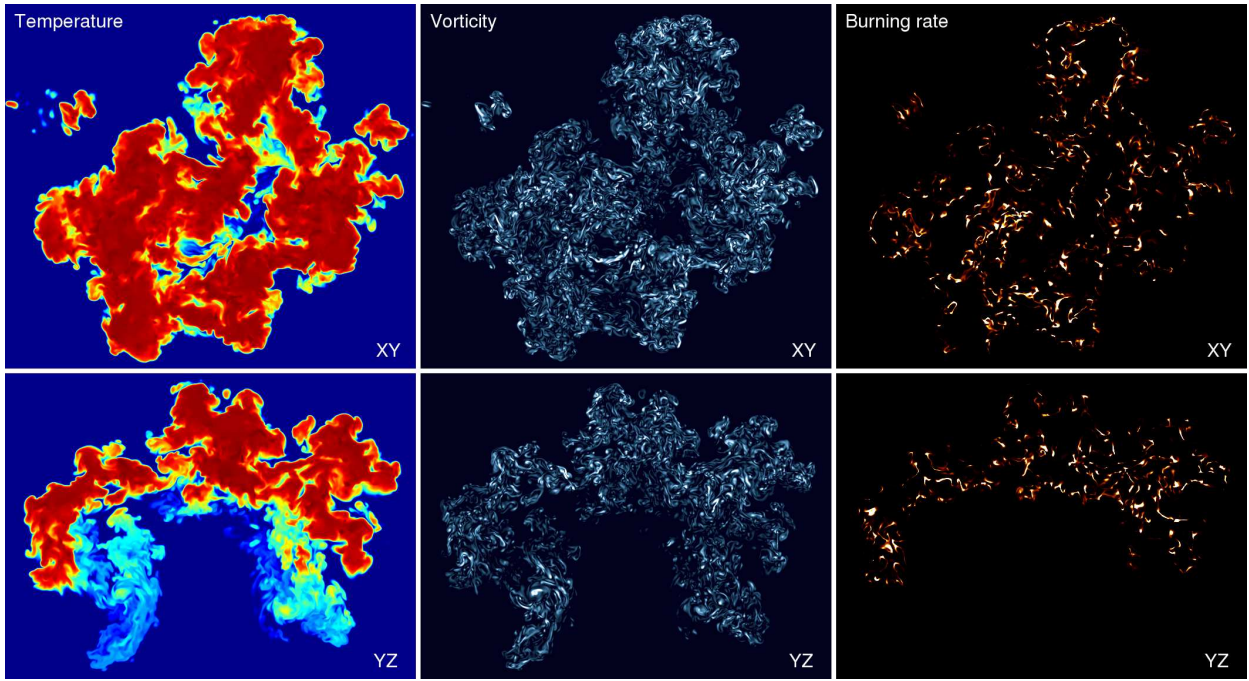}
\caption{Two-dimensional slices from the three-dimensional simulations
of a burning thermal in a uniform environment.  The top slices are horizontal,
and the bottom slices are vertical.  The panels left to right are temperature,
magnitude of vorticity, and burning rate.  All slices are shown at the same
scale and time $t=3.811$\,ms.}
\label{Fig:lateSlices}
\end{figure}

Two-dimensional slices through the data at $t=3.811$\,ms 
(figure \ref{Fig:lateSlices}) show the distributions of temperature, magnitude of 
vorticity, and burning, left-to-right, respectively.  By this stage of the evolution,
the thermal has become fully-developed, with burning and vorticity distributed
across the entire body of the thermal.  The temperature field shows that the thermal
still consists of regions of pure ash separated from pure fuel by relatively narrow 
mixing regions.  This differs from the distribution expected from an inert thermal 
(see figure \ref{Fig:earlySlices}), where mixing leads to a broader distribution of
temperature than the almost bimodal distribution seen in here, and is entirely due
to burning: the entrained fuel quickly burns to the ash temperature.
The distribution of burning across the thermal is of particular interest.  
Under these conditions, the burning occurs in pockets roughly evenly distributed
across the whole thermal.  This is indicative that the flame is dominated by the 
hydrodynamics, in particular, turbulent entrainment, rather than flame propagation
driven by thermal diffusion.  These observations support the assumptions that the
burning thermal can be considered to be at the constant ash density, and that
entrained fluid burns essentially instantaneously; the reactions are controlled
by turbulent mixing.

If the one-dimensional entrainment model developed in the previous section is appropriate, 
then once the thermal has become fully-developed, a self-similar evolution should be 
observed, where the relations predicted by equation \ref{eq:soln} should be satisfied,
from which the model constants $\alpha$, $\beta$, and virtual space-time origin 
$(z_{\mathrm{v}},t_{\mathrm{v}})$ can be identified.
To compare the three-dimensional simulations with the one-dimensional entrainment model, 
the thermal radius and height (and therefore velocity) were evaluated as a function of time.
This was achieved by integrating over the entire computational domain and classifying each
point in space as either inside or outside of the thermal based on a critical value of density.
Specifically, any point in space with density below 1$\times10^7$\,g\,cm$^{-3}$ was considered to be inside
the thermal, and otherwise outside the thermal.  This defines a volume $V$ over which the 
integrals of 1, $\rho$ and $\rho z$ were evaluated.  The first two quantities respectively define 
the volume and mass of the thermal, and the latter, when divided by the mass, defines
the vertical center of mass of the thermal, taken to be the thermal height.  The radius was 
taken to be that of a sphere with volume equal to that of the thermal, i.e. $b=(3V/4\pi)^{1/3}$.  

\begin{figure}
\centering
\plottwo{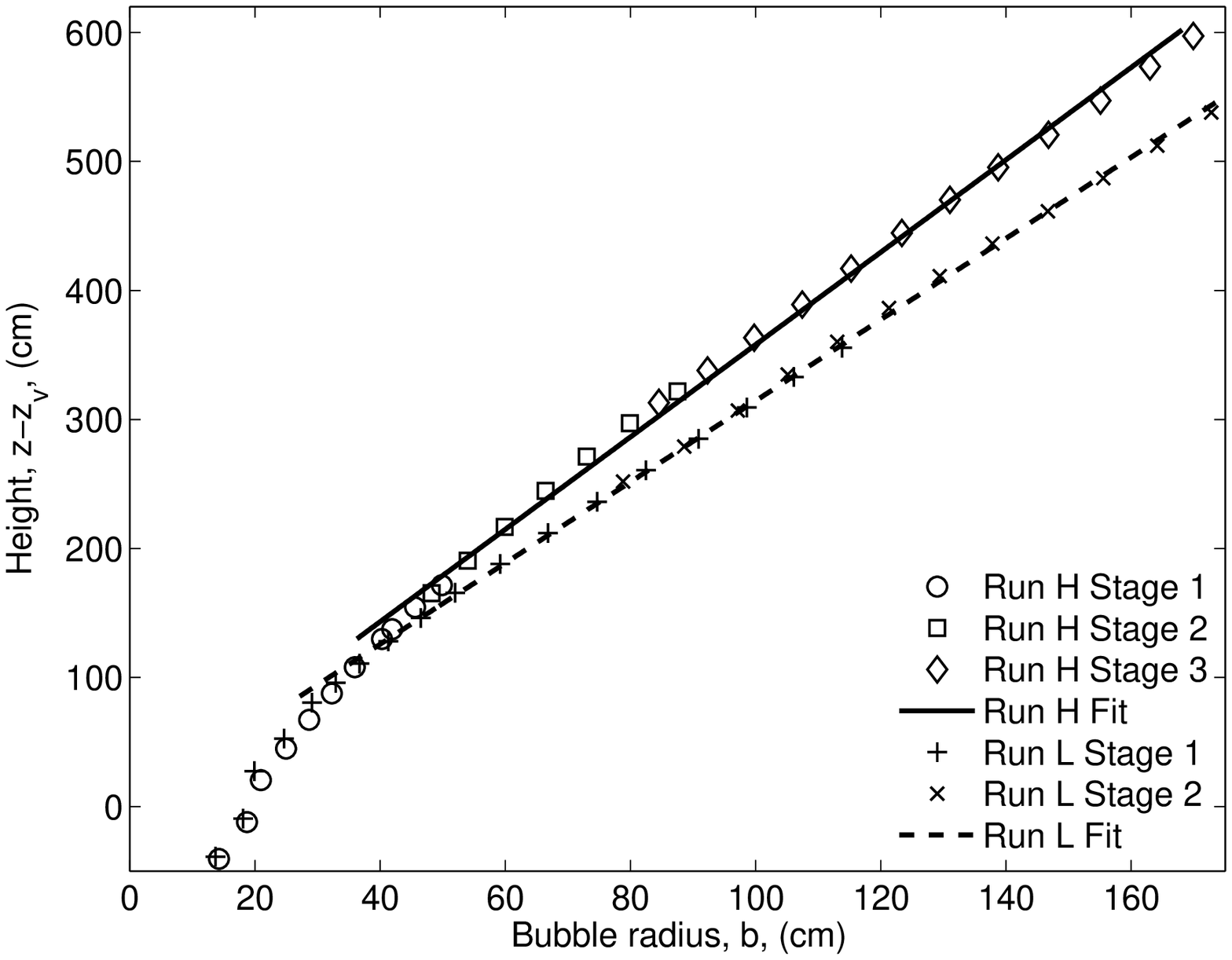}{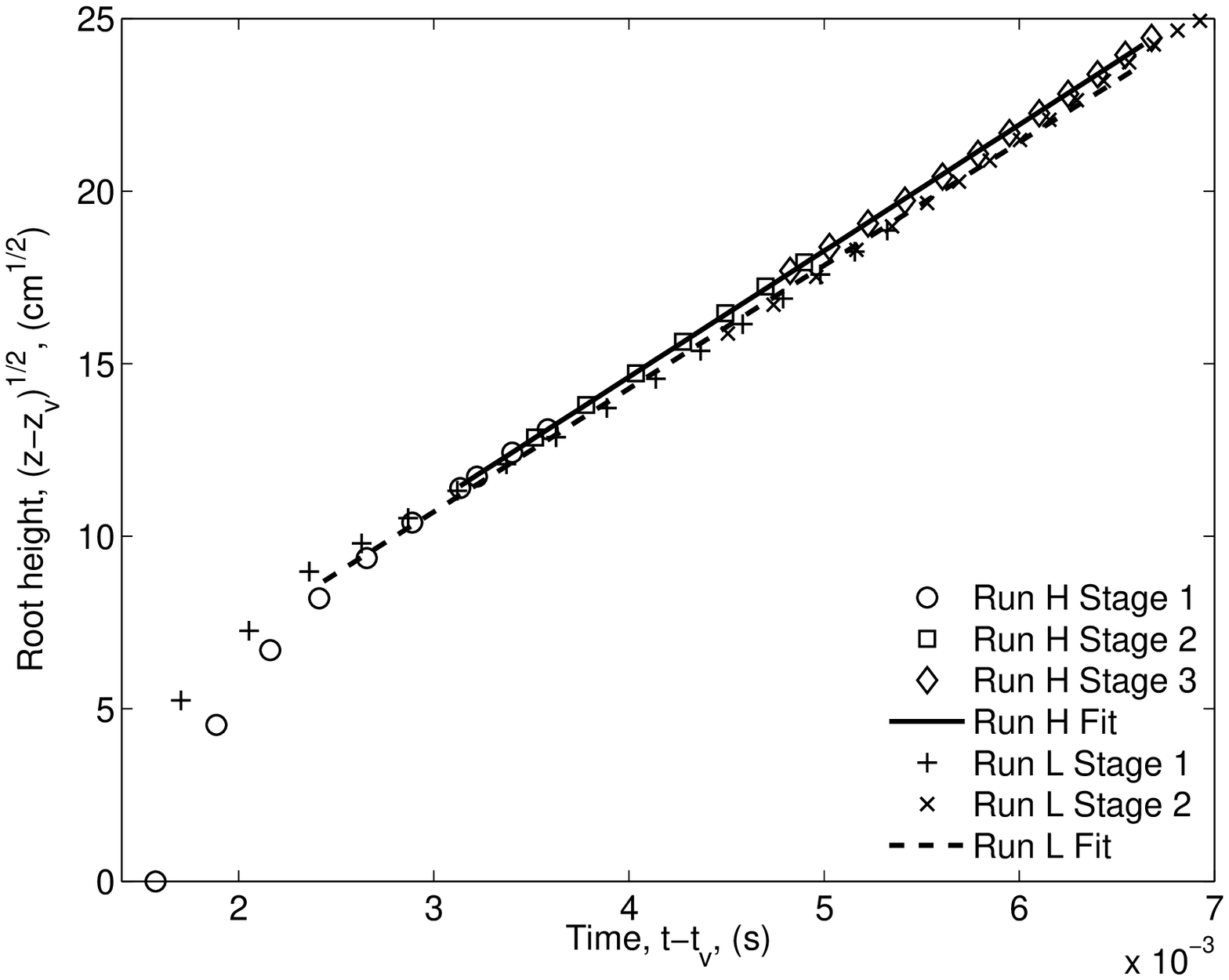}
\caption{(left) Normalized height of the thermal as a function of
  radius.  The thermal is predicted to prescribe a cone, shown by the
  solid line $ b = \sigma \alpha (z-z_{\mathrm{v}})$, where $\alpha$
  was measured through least-squares fitting to be approximately
  $0.17$ for run H and $0.19$ for run L.
(right) Evolution of the square root of the normalized height of the
  thermal.  After an initial transient, the rise of the thermal appears
  to rise as predicted; linear growth in
  $\sqrt{(z(t)-z_{\mathrm{v}})}=\sqrt{g^\prime/14}(t-t_{\mathrm{v}})$ demonstrates the expected
  quadratic growth.  The constant $\beta$ was determined through 
  least-squares best-fitting to be approximately $0.50$.}
\label{Fig:PowerLaws}
\end{figure}

To establish if the entrainment model is indeed appropriate, 
the data were compared with the predicted power-law evolution.
The entrainment coefficient $\alpha$ and virtual space origin $z_{\mathrm{v}}$ were first evaluated by 
performing a least-squares best-fit to the radius as a function of height, taking data after
the thermal has become fully-developed, i.e.\ $t\gtaprx2.05$\,ms.  Specifically, 
given a value of $\sigma=\rho_F/\rho_A$, $\alpha$ and $z_{\mathrm{v}}$ were found by a best-fit to
$b=\alpha\sigma(z-z_{\mathrm{v}})$.  The shape function $\beta$ and virtual time origin $t_{\mathrm{v}}$ were 
evaluated by a best-fit to the thermal height as a function of time,
i.e.\ $\sqrt{z-z_{\mathrm{v}}}=\sqrt{g^\prime/14}(t-t_{\mathrm{v}})$, where the value for $\beta$ is obtained
implicitly through $g^\prime$.  The best-fits were performed for both 
cases H and L.  The normalized height as a function of radius and square root of normalized 
height as a function of normalized time are presented in figure \ref{Fig:PowerLaws}.
The data from the simulations are presented as symbols (not all data points have been 
presented for clarity), where each kind of symbol corresponds to the various stages
of each simulation, note the overlap of data, confirming the validity of the choice
of continuation points.  In both figures, the agreement with the predicted power-law behavior
is compelling; height as a function of time is particularly robust ($\beta=0.50)$.  
There is a slight difference with resolution in the entrainment coefficient: $\alpha=0.17$ 
for case H, and $\alpha=0.19$ for case L, but this is well within the range observed in the 
laboratory experiments of \cite{Turner63} with constant acceleration (0.14-0.26), although 
we note that we have adjusted for the density ratio $\sigma$.  We will use the values 
$\alpha=0.17$ and $\beta=0.50$ for the full-star model.

\section{THERMAL BURNING IN A FULL STAR}

\subsection{Analysis}
\label{Sec:FullStarAnalysis}

We now consider the evolution of a burning thermal in a full-star, 
specifically to include the effect of the pressure gradient.
First, we require equations to describe the background stratification,
which we take to be spherically symmetric, we use $z$ to denote height 
to avoid confusion with the radius of the thermal.  For simplicity, 
we describe the equation of state solely by a relativistic 
degenerate electron pressure given by a polytrope of order $n=3$.  
Specifically, write pressure $p_F(z)$ as a function of density 
$\rho_F(z)$ according to 
\begin{equation}
p_F=K(Y_e\rho_F)^\gamma,
\quad\quad
\gamma=1+\frac{1}{n}=\frac{4}{3},
\label{eq:eos}
\end{equation}
where $Y_e=1/2$ and $K=1.244\times10^{15}$\,dyne\,cm$^{-2}$.
We assume the star is in hydrostatic equilibrium,
\begin{equation}
\frac{\D p_F}{\D z}=-\rho_F g,
\label{eq:hydrostatic}
\end{equation}
and $g(z)$ is given by Newton's law of universal gravitation
\begin{equation}
g(z)=\frac{4\pi G}{z^2}\int_0^z \rho_F \eta^2 \,\D\eta,
\label{eq:gravitation}
\end{equation}
where $G=6.67\times10^{-8}$\,cm$^3$\,g$^{-1}$\,s$^{-2}$.
Combining equations (\ref{eq:eos})-(\ref{eq:gravitation}), and 
assuming $\rho_F$ is finite at $z=0$, a single second-order 
non-linear ordinary differential equation can be obtained
for density as a function of height
\begin{equation}
\frac{1}{\rho_Fz^2}\frac{\D}{\D z}\left(\rho_F^{\gamma-2}z\frac{\D \rho_F}{\D z}\right)=
-\frac{4\pi G}{\gamma K Y_e^\gamma}.
\label{eq:lane-emden}
\end{equation}
In non-dimensional form, equation (\ref{eq:lane-emden}) is the 
Lane-Emden equation, which is known not to possess an
analytic solution for $n=3$.  This means a simple solution
of the kind found in the previous sections will not be 
forthcoming, and that numerical integration will be required to 
find a semi-analytic solution for a burning thermal in such an environment.

We now derive conservation equations for volume, mass and momentum
for a burning thermal of density $\rho$ rising in an ambient 
stratification of density $\rho_F$ satisfying (\ref{eq:lane-emden}).
A power law is used to describe the ash density $\rho_A$ as a function of 
fuel density $\rho_F$, given by $\rho_A=0.1168,\\rho_F^{1.091}$, 
where the constants were determined from a least-squares best-fit to
data from \cite{timmeswoosley1992} and \cite{Woosley2007}.
The Boussinesq approximation is
not appropriate due to large variations in density, which means 
the equations for volume and mass are no longer interchangeable.
As before, we assume that the thermal entrains fluid at a rate 
proportional to the rise speed, and that the entrained fluid mixes 
and burns instantaneously.  The entrainment coefficient is assumed to
be constant.  In this case, we also allow the thermal to burn independently
from entrainment by adding a flame speed $s$ to the rate of entrainment, 
i.e.\ the total rate of entrainment is taken to be $\alpha|u|+s$.  This is 
necessary due to the large turbulent flame speed used in the full-star simulation, 
and was not required in the previous section because the laminar flame 
speed is much slower.  This is particularly important when the thermal 
is stationary, but continues to burn outwards, which is not captured by 
the entrainment assumption.  The final assumption is that the thermal
rises isentropically and instantaneously expands and equilibrates 
to the ambient pressure at that height.  The change of specific 
volume $\tau$ associated with this expansion can be written as
\begin{equation}
\frac{\D \tau}{\D t}=-\frac{\tau}{\rho_F}\frac{\rho_F}{\D t}.
\label{Eq:FullStarExpansion}
\end{equation}
This means that conservation equations for volume, mass and
momentum can be written as
\begin{eqnarray}
\frac{\D}{\D t}\left(\frac{4}{3}\pi b^3\right) &=& 
4 \pi b^2 \sigma (\alpha |u| + s)
-\frac{4\pi b^3}{3\rho_F}\frac{\D \rho_F}{\D t},
\label{Eq:FullStarVol}\\
\frac{\D}{\D t}\left(\frac{4}{3}\pi b^3 \rho\right) &=& 
4 \pi b^2 \rho_F (\alpha |u| + s),
\label{Eq:FullStarMass}\\
\frac{\D}{\D t}\left(\frac{4}{3}\pi b^3 (k_v\rho_F+\rho) u\right) &=& 
\frac{4}{3} \pi b^3 \beta \left(\rho_F-\rho\right) g.
\label{Eq:FullStarMom}
\end{eqnarray}
The two terms on the right-hand side of equation (\ref{Eq:FullStarVol})
account for the change in volume of the thermal due to burning (as in 
the previous section) and isentropic expansion (following equation 
(\ref{Eq:FullStarExpansion})), respectively.  Equation (\ref{Eq:FullStarMass}) 
states that the total mass of the thermal increases by the mass of the 
entrained fuel.  Finally, equation (\ref{Eq:FullStarMom}), 
as before, is Newton's second law and describes the rate of change
of momentum due to buoyancy.  An equation for conservation of buoyancy
can be derived, but is not required.

Assuming that the rise of the thermal does not affect the background
stratification, and combining equations (\ref{eq:gravitation}), (\ref{eq:lane-emden}),
and (\ref{Eq:FullStarVol})-(\ref{Eq:FullStarMom}) with the expression for burning 
and $u=\D z/\D t$ (which is used as a change of variable from 
$z$ to $t$ where necessary), the entire system can be described by a system of six 
non-linear ordinary differential equations, which can be written in terms of
primitive variables as
\begin{eqnarray}
\frac{\D z}{\D t}&=&u,\label{Eq:PrimHeight}\\
\frac{\D g}{\D t}&=&4\pi G \rho_F u - \frac{2gu}{z},\label{Eq:PrimGrav}\\
\frac{\D \rho_F}{\D t}&=&-\frac{gu}{\gamma K Y_e^\gamma \rho_F^{\gamma-2}},\label{Eq:PrimFuel}\\
\frac{\D b}{\D t}&=&\sigma(\alpha|u|+s) + \frac{bgu}{3\gamma K Y_e^\gamma \rho_F^{\gamma-1}},\label{Eq:PrimRad}\\
\frac{\D \rho}{\D t}&=&\frac{3\sigma(\alpha|u|+s)}{b}(\rho_A-\rho)-\frac{\rho g u}{\gamma K Y_e^\gamma \rho_F^{\gamma-1}},\label{Eq:PrimRho}\\
\frac{\D u}{\D t}&=&g\beta\frac{\rho_F-\rho}{k_v\rho_F+\rho}-3\sigma(\alpha|u|+s)\frac{u}{b}\frac{k_v\rho_F+\rho_A}{k_v\rho_F+\rho}\label{Eq:PrimSpeed}.
\end{eqnarray}
Initial conditions are required for the thermal, specifically, density $\rho_0$, 
radius $b_0$, height $z_0$ and speed $u_0$.  Equations (\ref{Eq:PrimGrav}) 
and (\ref{Eq:PrimFuel}) can then be integrated numerically (in spatial form) 
from $g=0$ and $\rho_F=\rho_c=2.55\times10^9$\,g\,cm$^{-3}$ (the density of the star at its center) at $z=0$ 
out to $z=z_0$, giving values for $\rho_F$ and $g$ at t=0, completing the 
set of initial conditions.  The system (\ref{Eq:PrimHeight})-(\ref{Eq:PrimSpeed}) 
can then be integrated numerically to obtain the evolution of the thermal.

Again, an equation for the evolution of the Froude number can be derived, here
presented in spatial form assuming $u>0$,
\begin{equation}
\frac{\D}{\D z}\left(\Fr^2\right)=\frac{1}{b}\left[
2 - \sigma \alpha \Fr^2\left(6+\frac{k_v\rho_F+\rho_A}{k_v\rho_F+\rho}\right)\left(1+\frac{s}{\alpha u}\right)\right]
-\Fr^2\left(\frac{4\pi\rho_FG}{g} + \frac{\rho_Fg}{3\gamma p_F} - \frac{2}{z}\right).
\nonumber
\end{equation}
The first term on the right-hand side is similar to equation \ref{eq:SimpleFr}, and 
indeed reduces to the same with $s=0$ and $\rho=\rho_A$.  The second term
accounts for changes in the background stratification, and is a function of $z$ alone, i.e.\ it
does not depend on the thermal.  
Assuming $\Fr\sim1$, then the 
second term varies between approximately $10^{-9}$ and $10^{-7}$.  
Taking $1/b$ as an estimate, the first term varies between approximately $10^{-8}$ 
and $10^{-6}$.  This means the ratio of the two terms is about an order-of-magnitude, 
i.e.\ the second term is small but cannot be ignored.

\subsection{Numerical Approach}
\label{Sec:FullStarNumerics}

Full-star simulations of an off-center SNIa ignition were performed in three 
dimensions by \cite{Dong11} using the Eulerian, adaptive compressible code, 
CASTRO, described in \citet{CASTRO:2010}.  The hydrodynamics in CASTRO is 
based on the unsplit PPM methodology in \citet{ppmunsplit}.  The white dwarf 
was again modeled using the general equation of state described by 
\citet{timmes_swesty:2000}, which includes contributions from electrons, ions, 
and radiation.  For these calculations the effects of Coulomb corrections were 
included from the publicly available version of this EOS \citep{timmes_eos}.  
Although CASTRO includes a full gravitational potential solver,  for the 
simulations discussed here a simple monopole approximation to gravity was used. 

Adaptive mesh refinement in CASTRO follows the same approach as the 
low Mach number algorithm described in the previous section.
The same coarse-grained parallelization strategy is also used, but 
in conjunction with a fine-grained strategy using OpenMP within the nodes
to spawn threads that operate on a portion of the data associated with a
patch.  Using this strategy, we have been able to demonstrate 
scalability of CASTRO to more than 200K processors, see \citep{Scaling:2010}.

For the simulations presented here, the star was initialized using a one-dimensional
white dwarf model produced with the stellar evolution code Kepler 
\citep{weaver:1978}.  The initial composition was half $^{12}\mathrm{C}$ and 
half $^{16}\mathrm{O}$.  A flame was initialized by creating a bubble of 
hot ash at a temperature of $8.5 \times 10^9$\,K in pressure equilibrium with the
surrounding material, 20\,km in diameter located 30\,km off the center of the star.
The initial sphere was perturbed by a superposition of random spherical harmonics.

The flame was modeled using a thick-flame approximation (see \citet{colin2000})
with the burning time scale and thermal diffusion adjusted to produce
a flame that propagated at 50\,km\,s$^{-1}$. Composition of the ash was determined
by a 7 isotope tabular network that enforced nuclear statistical equilibrium
for sufficiently high temperatures.
The simulation was run with a base mesh of 16\,km with 4 levels of refinement
by a factor of 2 each. The region around the flame was always forced
to be refined to the finest level, so the flame zone had an effective resolution
of 1\,km.  Additional refinement was based on density so that the center
of the star was also refined to 1 km resolution with resolution dropping
to 4\,km at the outer edge of the star.

\subsection{Results}
\label{Sec:FullStarResults}

\begin{figure}
\centering
\plotone{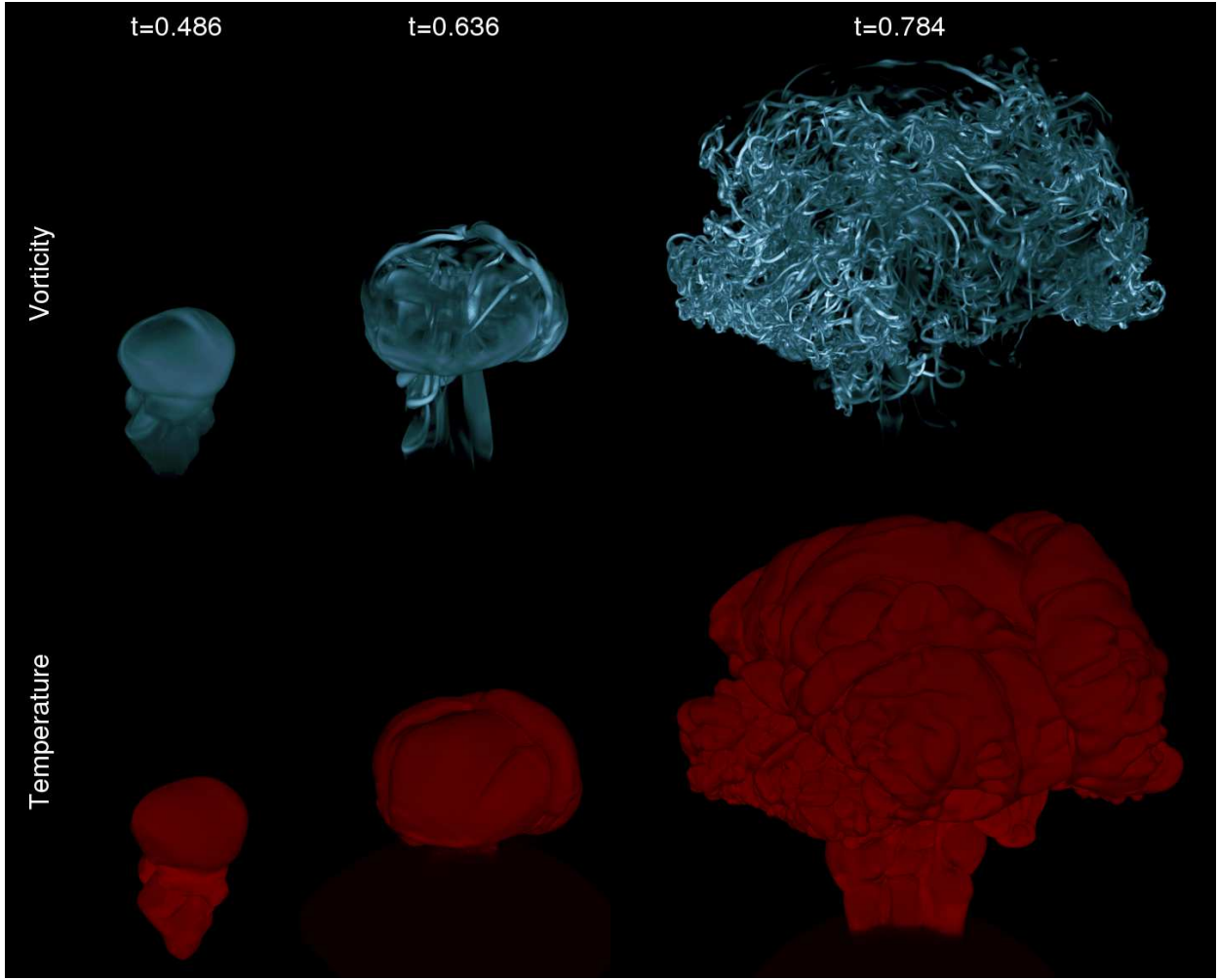}
\caption{Three-dimensional renderings of the evolution of a thermal
burning through a full star.  Simulation data taken from \cite{Dong11}.
Vorticity is shown in blue on the top row, and temperature in red on the bottom row.  Times are in seconds.
The full-star thermal follows a similar progression to the small-scale study; the thermal rolls up and forms
a toroidal structure, which breaks down due to secondary shear instabilities.  Further detail can be found
in \cite{Dong11}.
}
\label{Fig:FS3d}
\end{figure}

A series of three-dimensional renderings in figure \ref{Fig:FS3d} show the 
evolution of the thermal in a full star, which follows a similar progression as 
in the small-scale study.  The thermal was initiated with uniform burning over 
a perturbed sphere, which is very small in comparison with the star.  At early 
times, the thermal burns outwards without rising.  Buoyancy accelerates the 
thermal upwards, which rolls up into a bubble-like structure, which then breaks 
down due to secondary shear instabilities, with vortical structures distributed 
across the thermal.  The reader is referred to \cite{Dong11} for more detail.
The similarities between figure \ref{Fig:FS3d} and figures \ref{Fig:early3d} and
\ref{Fig:late3d} suggest that the entrainment-based modeling approach is 
appropriate for burning thermals in a full star.

We will make two comparisons of the one-dimensional model with the
three-dimensional simulation.  Firstly, for comparison A, we will set the flame speed $s$ to
zero, consistent with the previous section, and compare different values of
the entrainment coefficient $\alpha$, and also consider the case where
the volumetric expansion term is excluded from the model.  In this comparison, the 
one-dimensional model can only be expected to be applicable once the
thermal has become well-developed, which we will take to be $t\gtaprx0.1375$\,s.
Secondly, for comparison B, we will include the flame speed $s=50$\,km\,s$^{-1}$ to compare the full
one-dimensional model with the three-dimensional simulations, and include
the cases with $s=0$ and $\alpha=0$ for comparison.  In this case, it was found
that the one-dimensional model was applicable from an earlier time, specifically
$t\gtaprx0.0413$\,s, where it was noted that the vertical speed was zero so the flame burns 
outwards in a way not described by entrainment alone.  All of the one-dimensional
cases cannot be expected to be applicable at late times because the expansion
of the star is not included in the model, i.e.\ $t\gtaprx0.8175$\,s.

To integrate the one-dimensional system of equations, the standard fourth-order 
Runge-Kutta integrator included with MATLAB was used.  Initial conditions for
$t_0$, $z_0$, $u_0$, $b_0$ and $\rho_0$ were taken from the three-dimensional 
simulation data, which were evaluated by integrating the data in the same way 
as the previous section using $X_C=0.4$ as the threshold.
Time points $t=0.1375$\,s and $0.0413$\,s were used, respectively for comparisons A and B. 
Values for $\rho_F$ and $g$ at $z=z_0$ were found by integrating numerically the 
spatial form of equations (\ref{Eq:PrimGrav}) and (\ref{Eq:PrimFuel}) from 
$z=0$, $g=0$ and $\rho_F=\rho_c$ to $z=z_0$.  Equations (\ref{Eq:PrimHeight})-(\ref{Eq:PrimSpeed})
were then integrated numerically forward in time until $z(t)\approx1500$\,km.

\begin{figure}
\centering
\plottwo{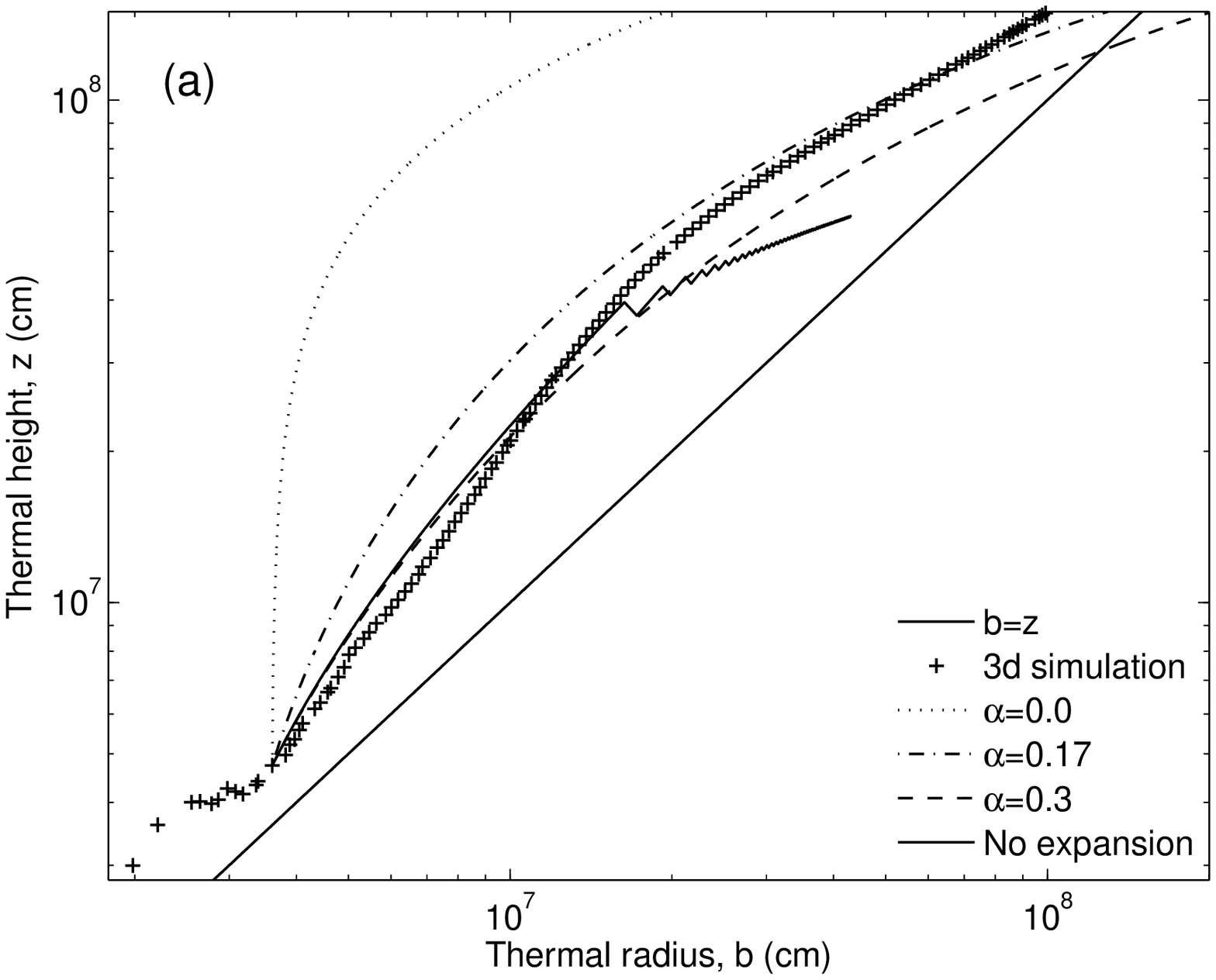}{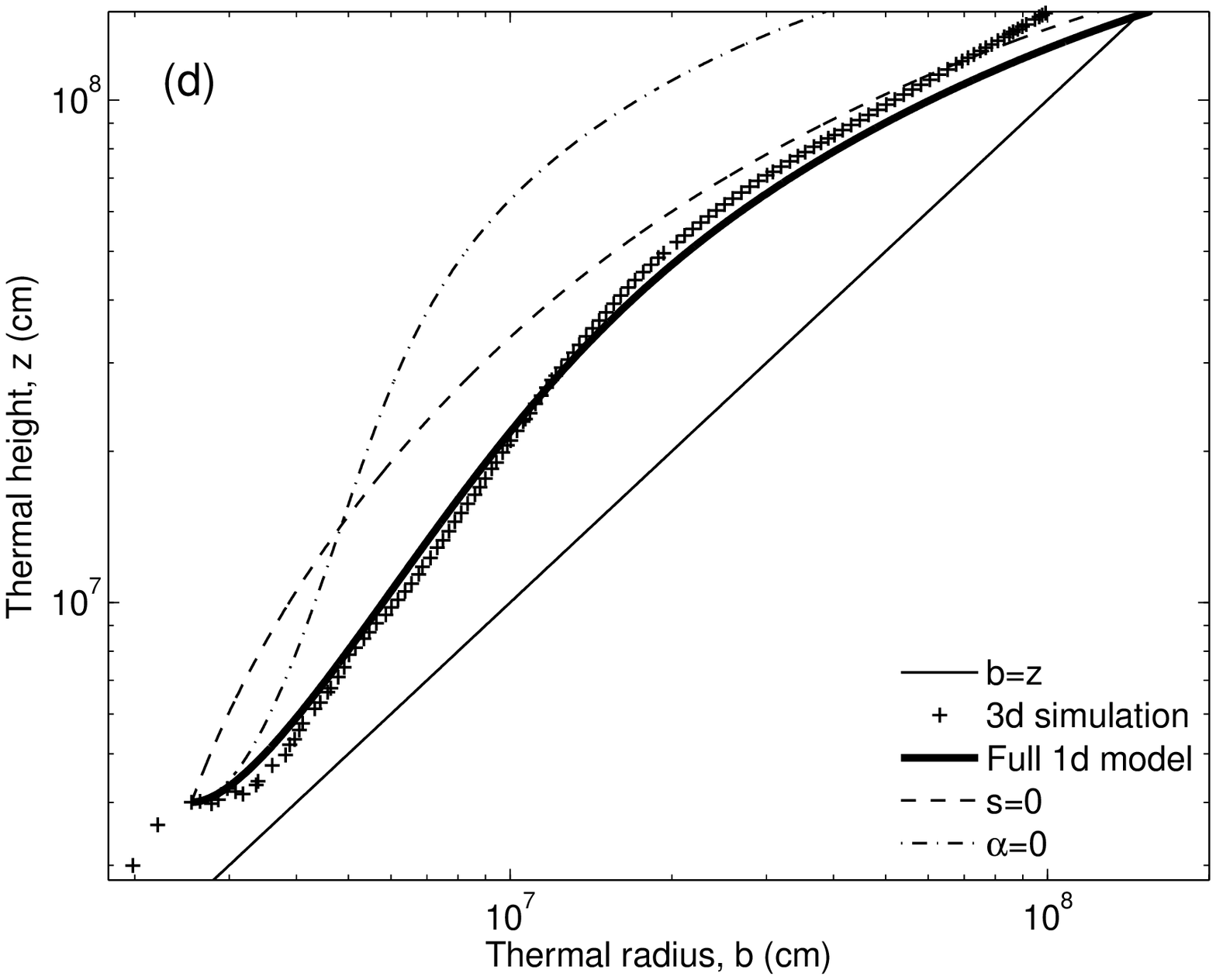}
\plottwo{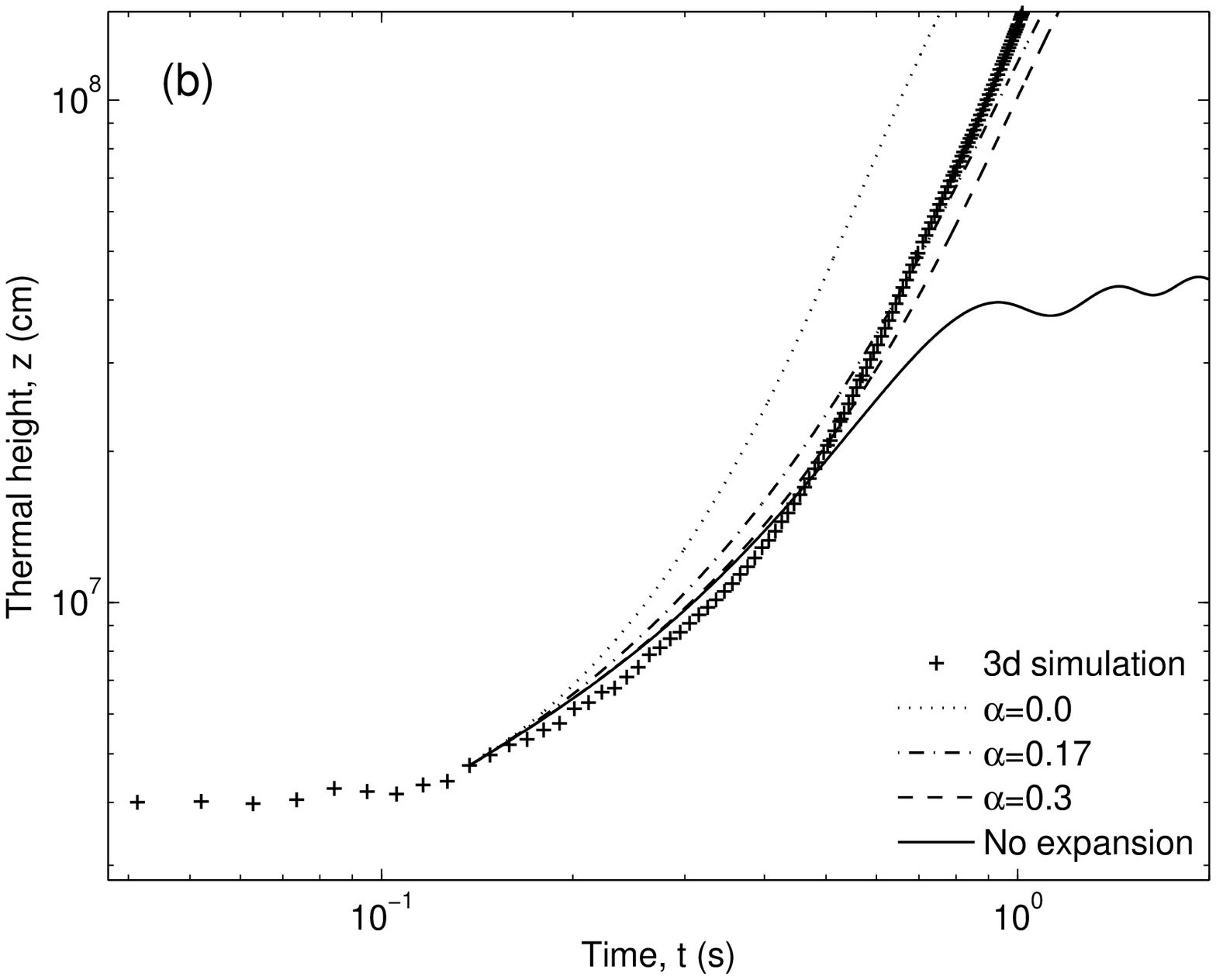}{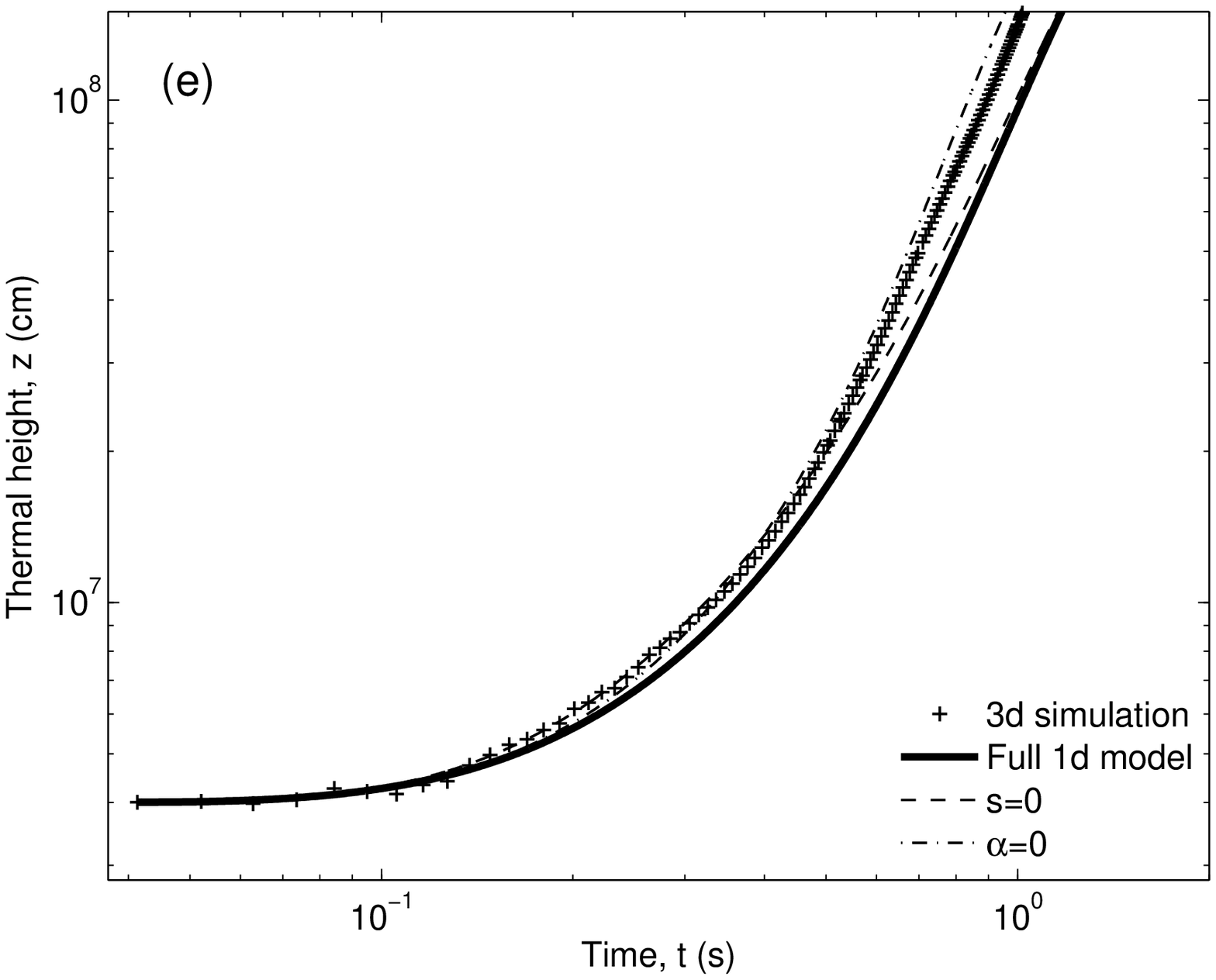}
\plottwo{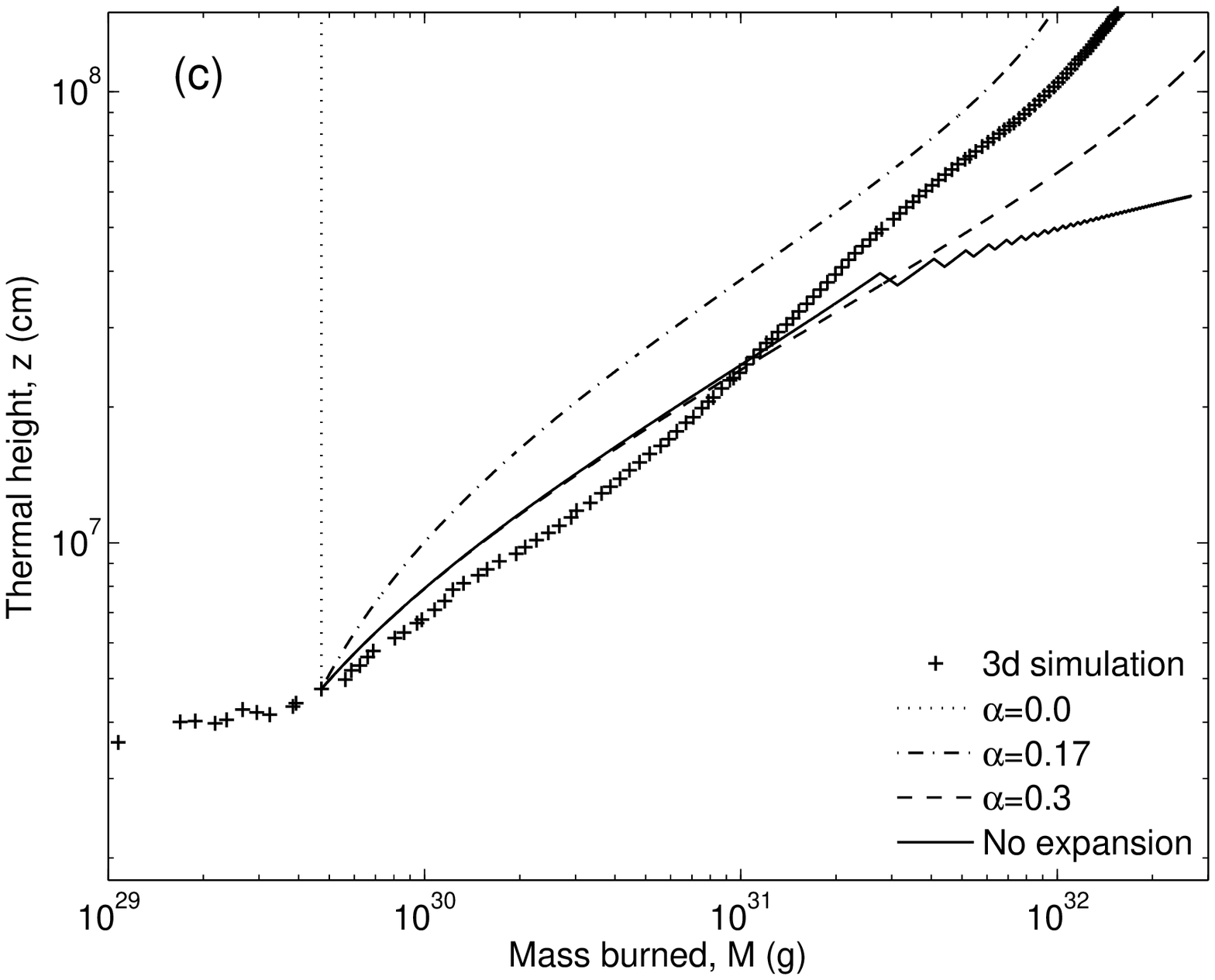}{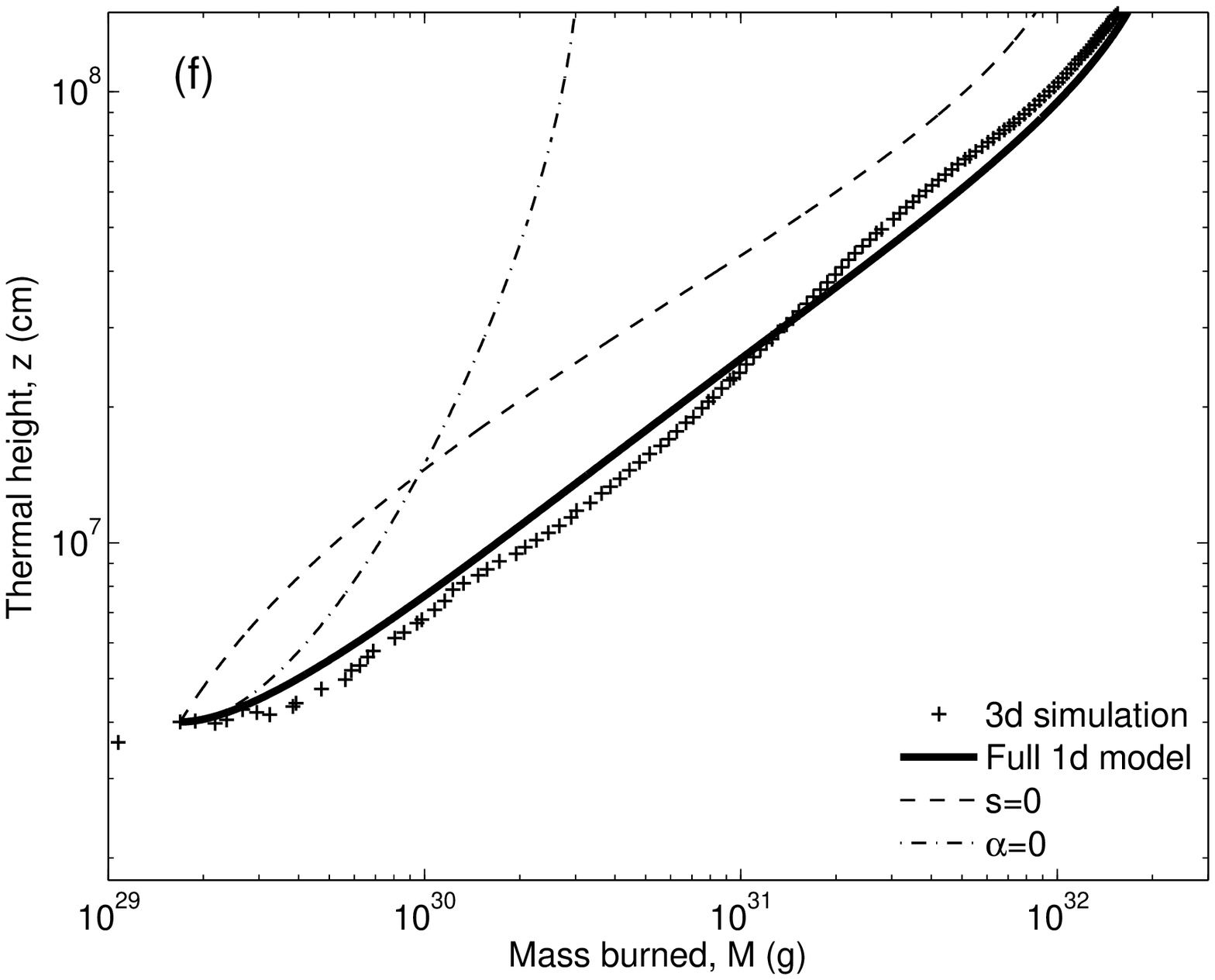}
\caption{Comparisons A (a-c) and B (d-f) of the three-dimensional simulation
of a burning thermal in a full star with the one-dimensional model with
different combinations of model constants.  Comparison A compares the effect 
of the entrainment coefficient $\alpha$ and the volumetric expansion term with
the flame speed $s$ set to zero, starting after the thermal has become more
well-developed, $t_0=0.1375$\,s.  Comparison B compares the full one-dimensional
model with and without entrainment and the flame speed included, starting 
from an earlier time $t=0.0413$\,s.}
\label{Fig:fsModel}
\end{figure}

The height of the thermal is plotted against radius, time and total mass burned 
for comparison A in figures \ref{Fig:fsModel}(a), (b) and (c), respectively.
The plus symbols denote the three-dimensional simulation data.  The dotted curve
denotes the thermal rising without entrainment ($\alpha=0$), which rises due to
buoyancy (faster than in the other cases) and expands due to the change in background 
stratification only (therefore, much less than in the other cases).  Naturally, 
the mass burned is constant.  The dash-dotted and dashed curves denote $\alpha=0.17$ 
(the value found in the previous section) and $0.3$ (for comparison), respectively.  
The inclusion of entrainment greatly increases the expansion of the thermal (as
expected), increasingly so with $\alpha$, but even with $\alpha=0.3$ the
early time evolution does not expand as much with height or burn as much fuel as the 
three-dimensional data.  The later times expand and burn more than the 
three-dimensional simulation, such that the thermal radius is greater than its height,
meaning that it has burned back through the center of the star (see figure 
\ref{Fig:fsModel}(a) where the dashed curves are to the right of the solid
straight line $b=z$).  Both entraining cases match the time-scale more 
closely, but still appear to rise more quickly than the three-dimensional data.
The solid curve denotes $\alpha=0.3$ without the volumetric expansion term.
In this case, the lack of expansion as the background conditions change means that
the buoyancy of the thermal decreases, it becomes neutrally buoyant,
and the thermal rise speed drops.  The thermal then overshoots the point of neutral
buoyancy, it becomes negatively buoyant, and the rise speed becomes negative.
The thermal continues to oscillate around a slowly increasing height.

The height of the thermal is again plotted against radius, time and total mass burned
for comparison B in figures \ref{Fig:fsModel}(d), (e) and (f), respectively.
Again, the plus symbols denote the three-dimensional data.  In this case, the
initial condition is taken earlier than comparison A, $t=0.0413$\,s compared with
$t=0.1375$\,s.  The full one-dimensional model, with $\alpha=0.17$ and $s=50$\,km\,s$^{-1}$, 
is shown by the thick solid curve.   This is compared with two other cases
first with the flame speed turned off $s=0$ (dashed curve), and second with entrainment 
turned off $\alpha=0$ (dash-dotted curve).  At early times, the thermal is stationary (and
therefore not entraining), but burns outwards due to the flame speed $s$, so the full
model and $\alpha=0$ case agree for a short time.  As the thermal begins to rise, 
entrainment becomes more dominant and the $\alpha=0$ curve does not expand sufficiently
rapidly with height.  The flame speed term remains important though, as closer agreement
is found here than was found in comparison A with $\alpha=0.17$ and $s=0$.
The $s=0$ case in comparison B does not expand sufficiently quickly
at early times, because the thermal is initially stationary, and only begins to
entrain as it begins to rise due to buoyancy, which is why comparison A was made
with the later initial start time $t_0=0.1375$.  The full one-dimensional model
appears to follow the expansion with height of the thermal closely, even as
it turns over above $z\gtaprx500$\,km, but it still over-predicts the spread above
$z\gtaprx1000$\,km, although this is not surprising since the expansion of the
star is going to be important by this stage.  The full-star model predicts a slightly
slower rise speed than in the three-dimensional simulation, but provides the best 
prediction of the total mass burned.  At a height of 1500\,km (the extent of the 
one-dimensional model, beyond which the neglected expansion of the star cannot be 
ignored), the mass burned in the model is approximately 
$0.0840M_\odot$, which is only about 6\% more than at the same height in the three-dimensional 
simulation, $0.0794M_\odot$.  We emphasize that this is a semi-analytic result in the
sense that the error in the one-dimensional numerical solution is essentially negligible.

\subsection{Effect of initial conditions}

To examine the effect of initial conditions on the evolution of the thermal,
a series of calculations were performed using the full one-dimensional model.
This emphasizes the benefit of the one-dimensional model, as this study
is infeasible in three dimensions.  The quantities that were varied were 
the initial height ($z_0=16$, $40$, and $100$\,km), 
the initial radius ($b_0=10.2$, $25.5$, and $63.8$\,km), 
and the turbulent flame speed ($s=10$ and $100$\,km\,s$^{-1}$).
Note that the cases where $b_0>z_0$ were not used.  These initial heights and 
radii are consistent with \cite{Zin11}, and the turbulent flame speeds are 
a factor of five slower and two faster that in the previous section, respectively.  We note 
$100$\,km\,s$^{-1}$ is a little high, but these values are intended to provide 
lower and upper bounds.  The initial rise velocity was taken to be zero, and the 
initial mass burned taken to be the same value as in the previous section.

\begin{figure}
\centering
\plottwo{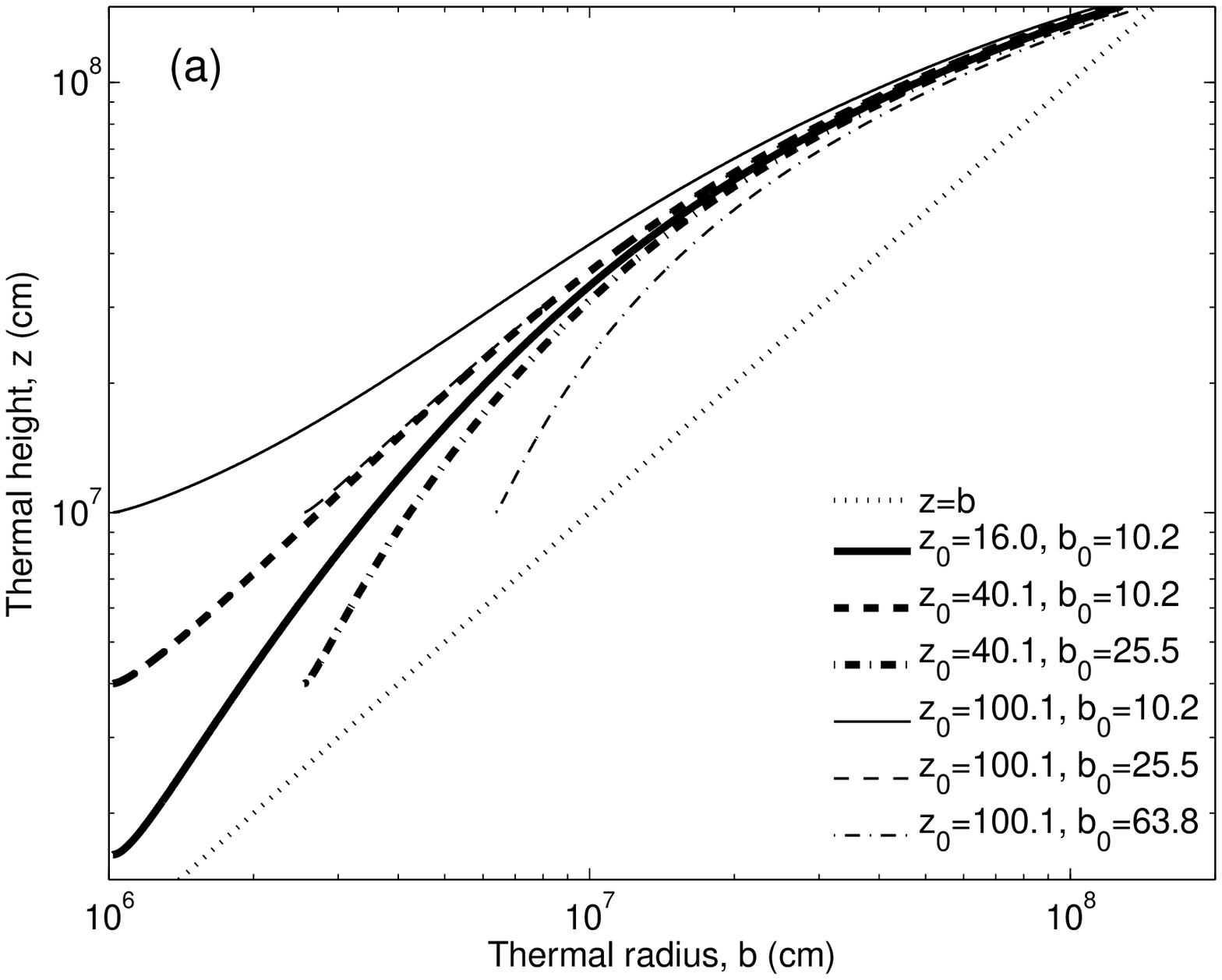}{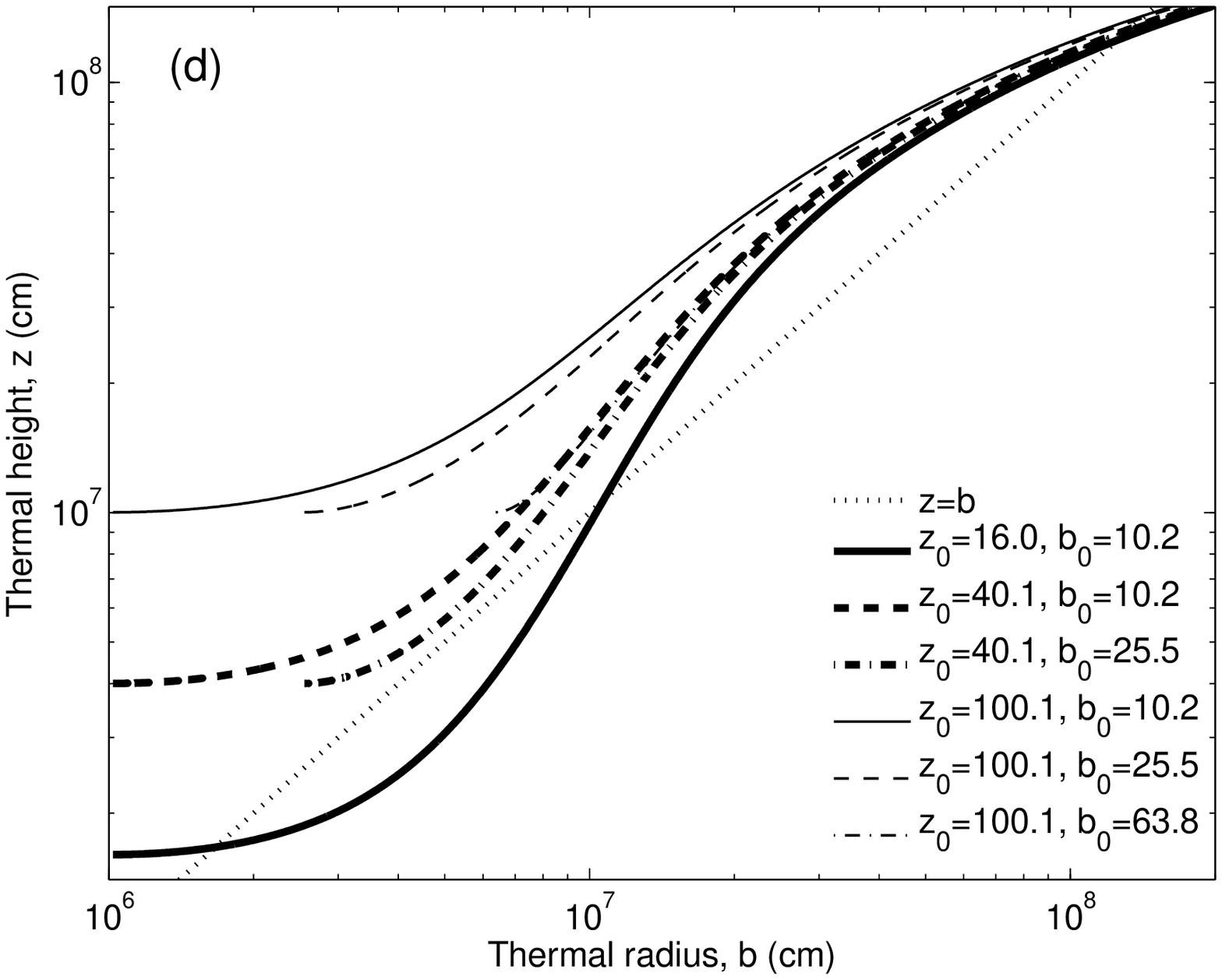}
\plottwo{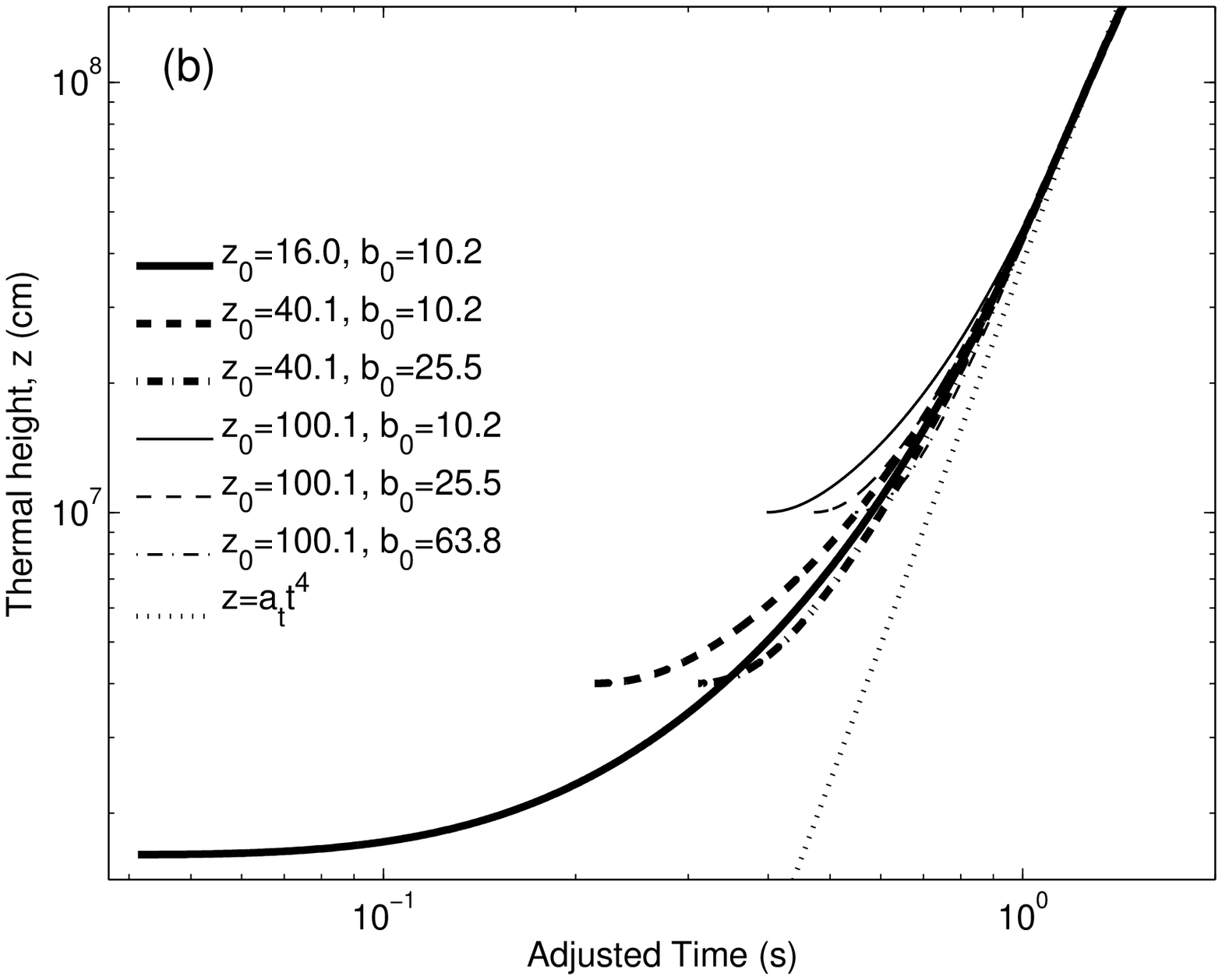}{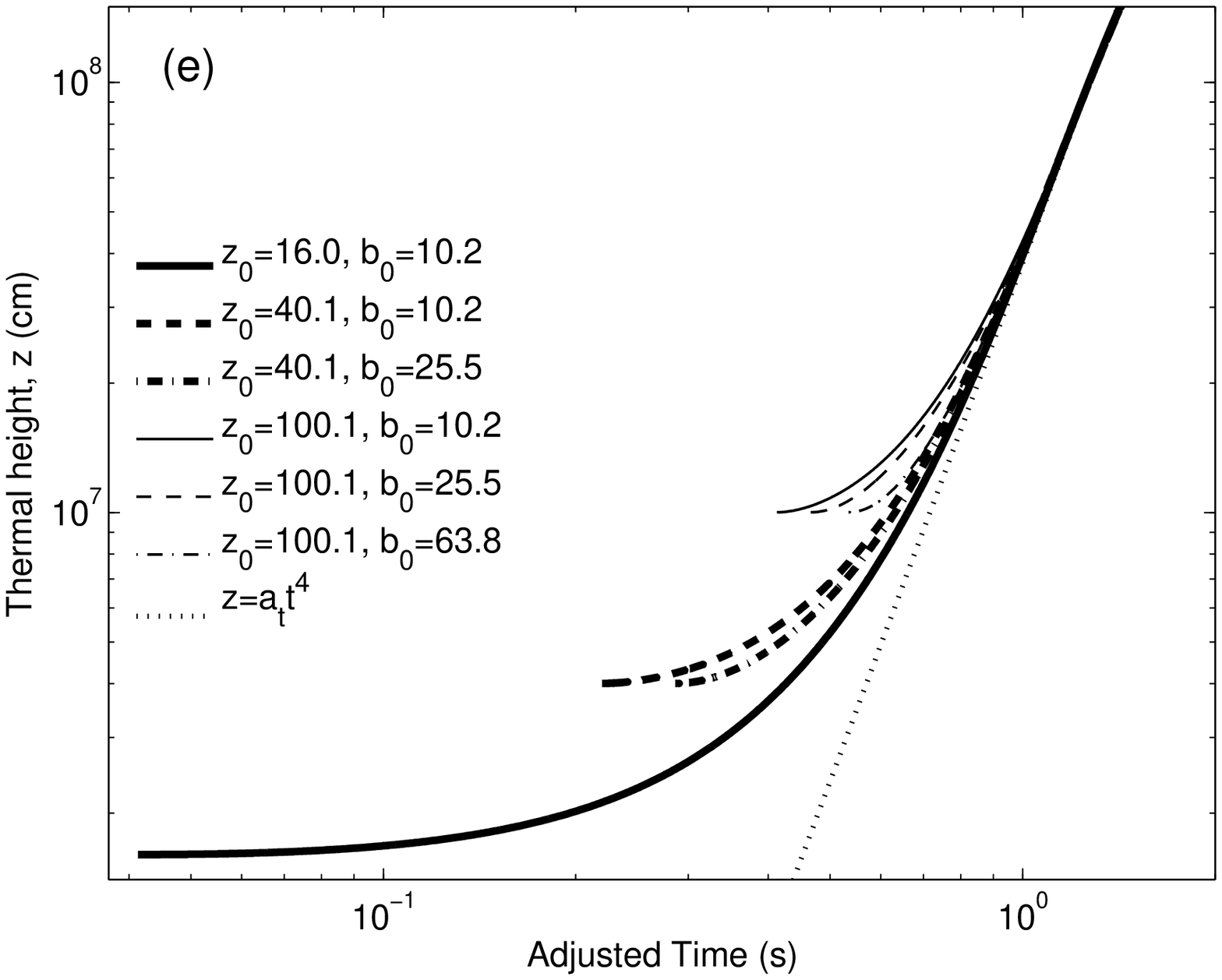}
\plottwo{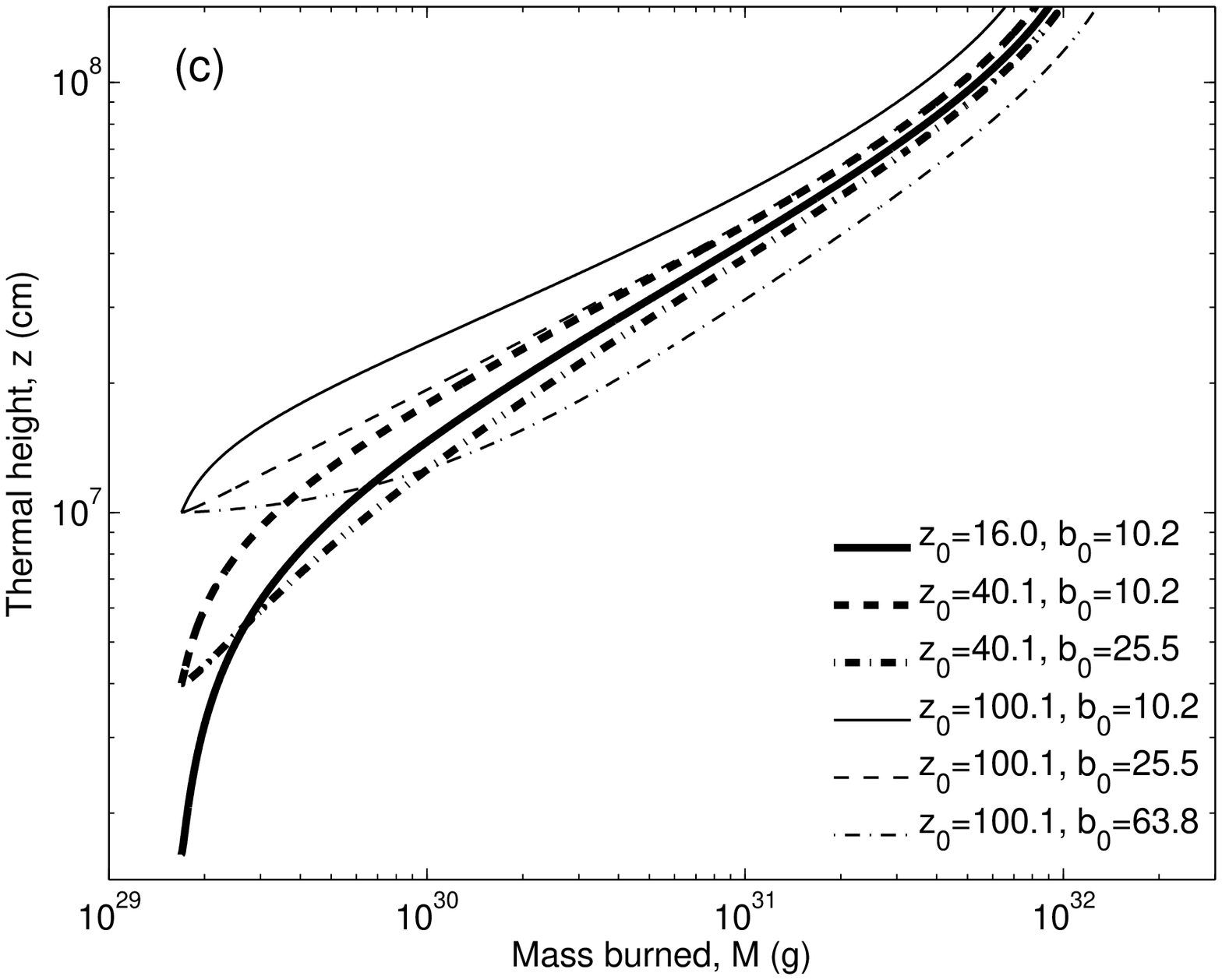}{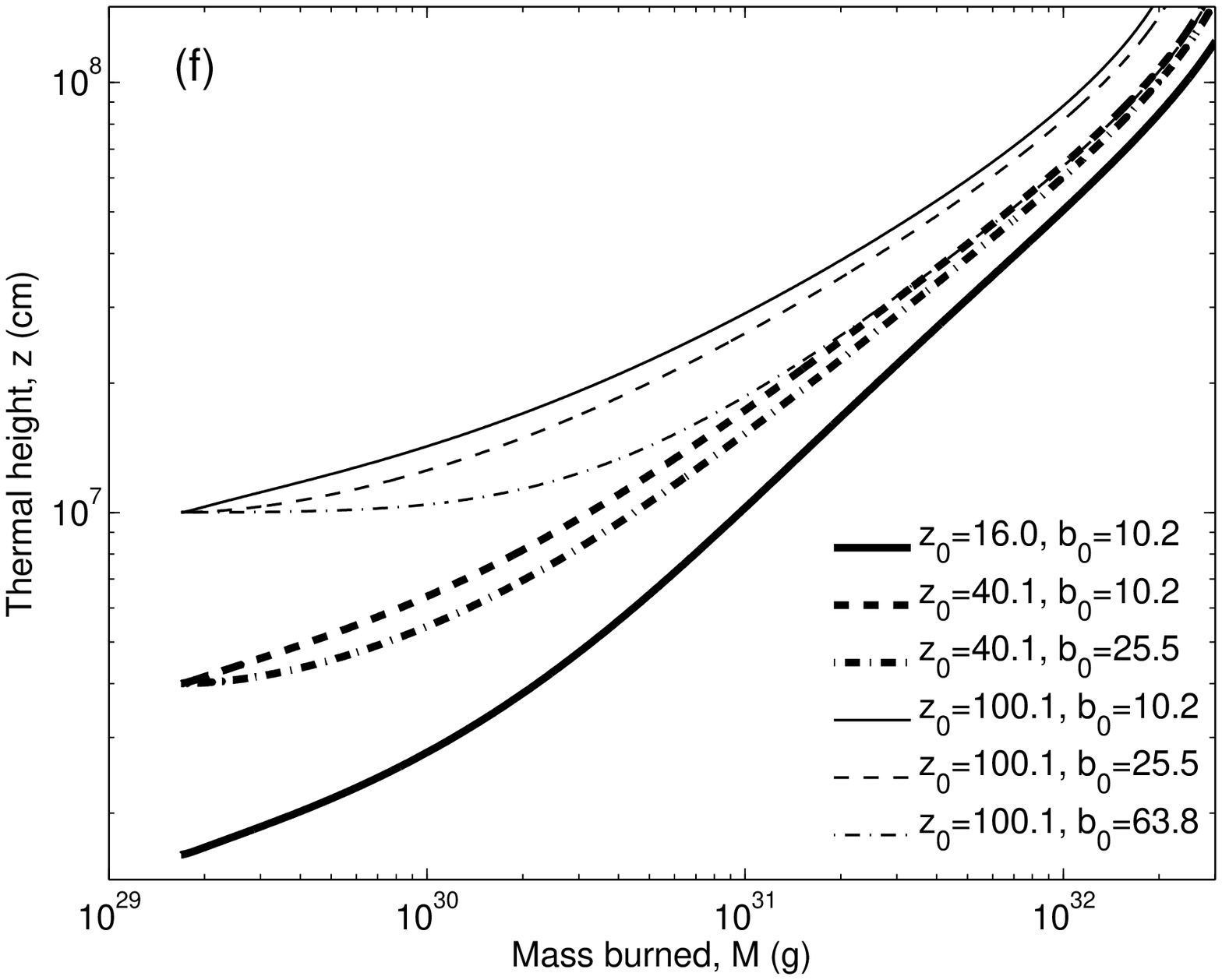}
\caption{Effect of initial conditions on the development of the thermal.  
(a)-(c) has $s=10$ km s$^{-1}$, and (d)-(f) has $s=100$ km s$^{-1}$.
The higher flame speed has a pronounced effect at early times as the thermal burns outwards 
before rising, giving the effect of a larger ignition kernel.  In general, more mass is 
burned for larger initial radii, smaller initial heights, and higher flame speed due to the 
early growth.}
\label{Fig:ICs}
\end{figure}

The evolution of the thermal height as functions of thermal radius, time, and mass burned
are shown in figure \ref{Fig:ICs}, where (a)-(c) correspond to $s=10$\,km\,s$^{-1}$, 
(d)-(f) are for $s=100$\,km\,s$^{-1}$, and the time has been shifted in (b) and (e) so that
the data coincide at a height of 800\,km.  In each case, the thermal initially burns outwards,
accelerates upwards, and continues to grow due to burning, entrainment and expansion.
The higher flame speed has a pronounced effect at early times as the thermal burns outwards 
before rising, giving the effect of a larger ignition kernel.  For the case where $z_0=16.0$, 
$b_0=10.2$, $s=100$, the thermal actually burns through the center of the star ($b>z$) and 
would likely result in a central ignition, which is not recovered by the model as it is 
assumed to be a point thermal.  For large $z$ the radii of the thermals tend towards similar 
values, becoming within a factor of about 2 of one another.  However, since mass burned depends on radius cubed, 
this results in values that differ by an order of magnitude (again note that the cases with the 
largest values of mass burned will have more likely resulted in central ignitions).  In general, 
more mass is burned for larger initial radii, smaller initial heights, and higher flame speed 
due to the early growth.  The thermal height against adjusted time appears to be insensitive 
to the initial conditions for large $z$.

\begin{figure}
\centering
\plottwo{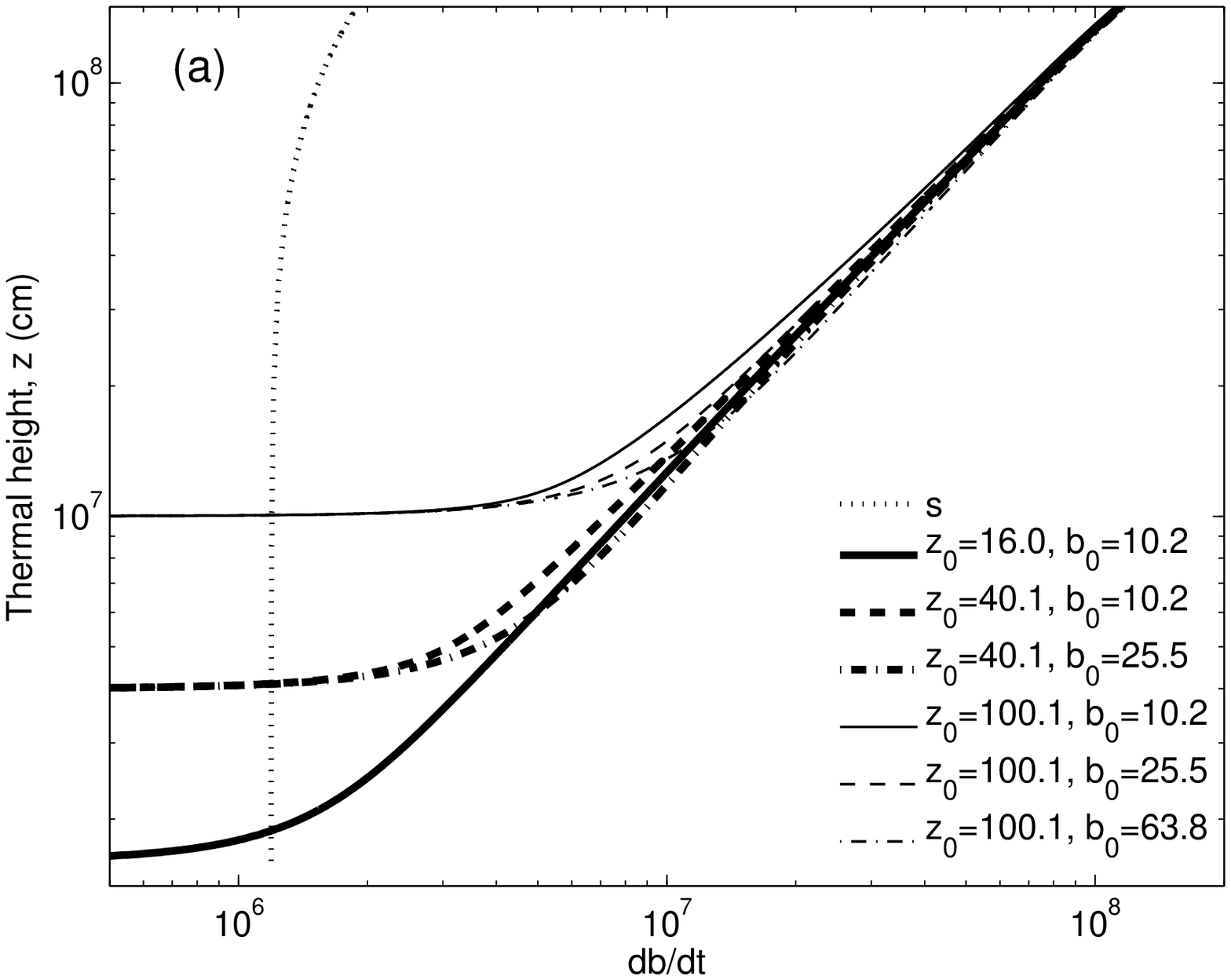}{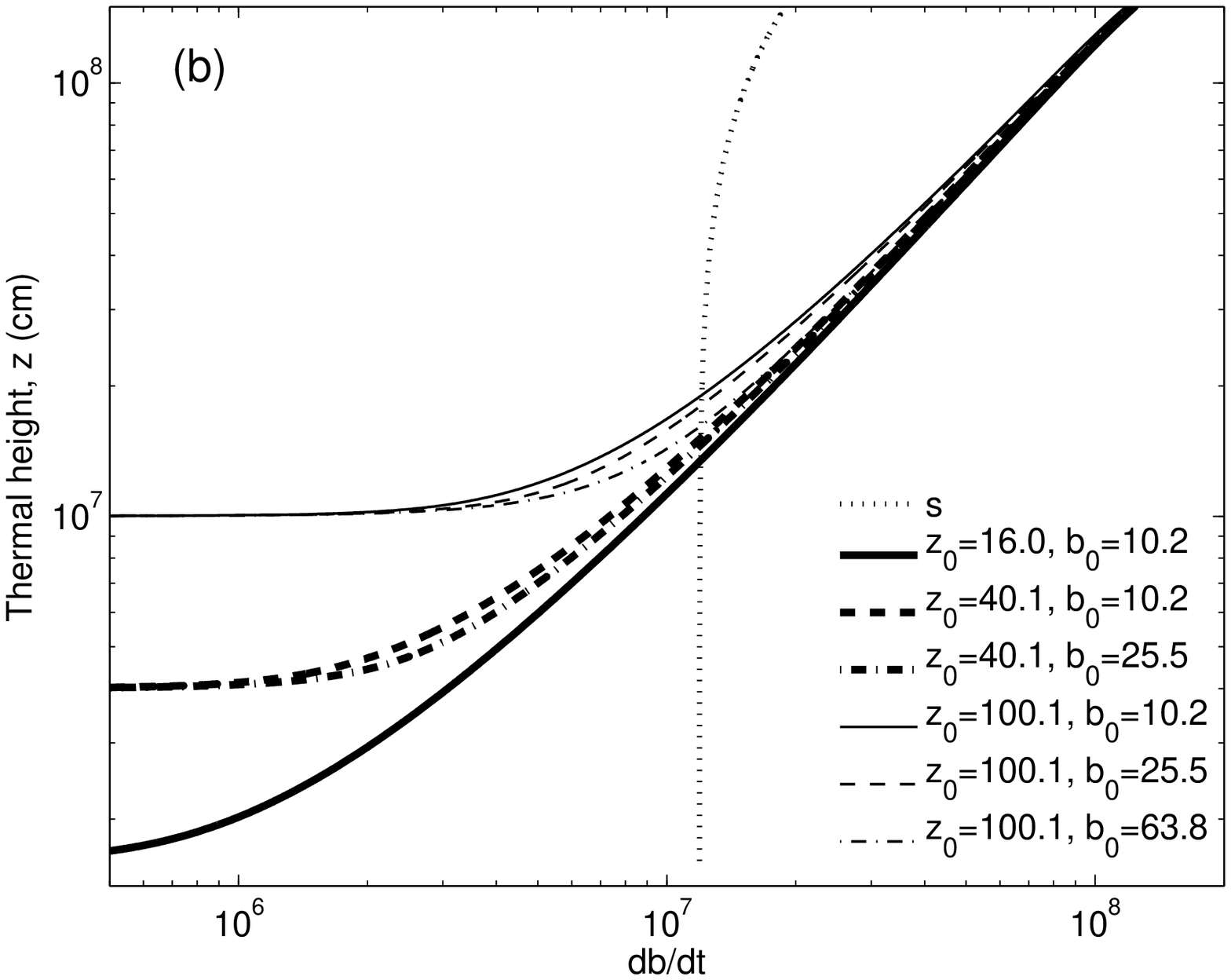}
\caption{Temporal rate of change of thermal radius due to entrainment ($\D b/\D t=\sigma \alpha |u|$) 
compared against turbulent flame speed ($\D b/\D t=\sigma s$) for (a) $s=10$\,km\,s$^{-1}$ and 
(b) $s=100$\,km\,s$^{-1}$.  The effect of flame speed is a function of height alone (through $\sigma$) 
and is shown by the near vertical dotted line.  The other lines correspond to the different initial 
conditions as in figure \ref{Fig:ICs}.  Once the thermal has become well-developed, the turbulent 
flame speed plays a diminished role, and the evolution is dominated by the large-scale hydrodynamics 
responsible for entrainment.
Note that 100\,km\,s$^{-1}$ is somewhat high, and may not be achievable, so 10\,km\,s$^{-1}$
may be more a relevant comparison.}
\label{Fig:UvsSF}
\end{figure}

Finally, figure \ref{Fig:UvsSF} compares the rate of entrainment against the turbulent flame
speed for each of the cases.  This suggests that once the thermal has become well-developed,
i.e.\ during the buoyant convection stage, the turbulent flame speed plays a diminished role, and the
evolution is dominated by the large-scale hydrodynamics responsible for entrainment. 
We note that the flame speed was included in the model to match the large-scale three-dimensional 
simulation which used an assumed turbulent flame speed, so it is possible that the flame speed term 
is overstated here and it should be even less important during the buoyant convection phase.
Furthermore, we emphasize that 100\,km\,s$^{-1}$ is somewhat high, and may not be achievable.
However, as noted above, the flame speed appears to be important at the early stages of the 
development, which suggests future studies should focus on the initial transition to turbulence 
with as detailed a flame model as possible.

\section{CONCLUSIONS}
\label{sec:conclusions}

The evolution of an off-center ignition in a Type Ia supernova explosion progresses
through multiple stages.  There is an initial transient as the ignition kernel begins 
to rise and transitions to turbulent evolution.  This is followed by the propagation 
of the bubble from the center to the periphery of the star by buoyant convection.
The final stage results from the interaction of the bubble with the boundary of the 
star, the outcome of which depends critically on the mass burned by the bubble, which 
affects the expansion of the star.  In this paper we deal with the intermediate stage 
of turbulent buoyant convection, which we have demonstrated can be well-described by 
a one-dimensional entrainment model that we have developed.  The model is based on the 
entrainment assumption of \cite{Morton56} and has been extended to account for both 
burning and for the change in background stratification of a full star.  To account for 
burning, the Boussinesq approximation was dropped and it was assumed that the 
entrained fluid was not only mixed, but also burned instantaneously.  The density ratio 
of fuel to ash was then used to account for burning in the volume conservation equation.
To account for the background stratification, a polytropic equation of state was used
in conjunction with hydrostatic equilibrium, and again the thermal was assumed to
equilibrate isentropically and instantaneously to the ambient conditions.

The model was first compared with a high-resolution three-dimensional simulation 
of a burning thermal in a uniform environment.  The burning was observed to occur
in pockets distributed evenly across the interior of the thermal, suggesting that the 
large-scale hydrodynamics, not the local flame speed, 
controls the rate of mixing and burning.  
In this simple scenario, the model can be solved analytically to provide power laws 
for the evolution in the asymptotic self-similar regime.  The power laws were in 
excellent agreement with the three-dimensional data and provided model constants 
that were carried forward to the full-star model.  

The Lane-Emden equation is known not to admit a power-law solution for polytrope order 
$n=3$, and so the full-star model had to be integrated numerically.  The expansion of 
the rising thermal was captured by assuming that it instantaneously equilibrated to 
the ambient pressure, giving rise to another term in the conservation of volume 
equation.  It was found that the inclusion of an extra term in the entrainment was 
required to account for the large model flame speed in the full-star simulation.  
However, once all of the components were assembled, the one-dimensional model provided 
compelling agreement with the three-dimensional simulation.  This means that
each component of the model (entrainment, expansion, background stratification, and turbulent flame speed)
is important for accurately predicting the evolution of the thermal.
The success of the entrainment model in predicting the evolution of the burning 
thermal in a full star again suggests that the dynamics, not the local flame 
speed, controls the rate of mixing and burning.

There are many aspects that have been neglected, but in particular, the expansion of 
the star has not been included in the model; hence the model can only be expected to
be valid at relatively early times when this expansion is not important.  Also, there 
are other terms in the equation of state that may be important at later times.

The one-dimensional model is computationally inexpensive, especially compared with
the three-dimensional simulations, which permitted a study of the effect of initial 
conditions that would have been infeasible in three dimensions.
The data demonstrate that there is some sensitivity to the initial conditions.  
More mass was burned with a larger initial radius and lower initial height, and
a higher flame speed gave the effect of a larger initial kernel resulting in more mass being 
burned.  This means that the very early stages of development are
of critical importance in predicting the overall evolution of the thermal.
In particular, the initial position and size along with how the early-time flow 
transitions to turbulence and self-similar evolution will all be extremely important to
the total mass burned.  Furthermore, the present study does not take into account
the background turbulence that results from the large-scale convection present in the star,
which may also have significance for the size and position of ignition kernels that can
survive.

\acknowledgements

The authors would like to thank Mike Zingale for his useful discussions.
Support for A.~J.~A.\ was provided by the Applied Mathematics Program of the
Office of Advanced Scientific Computing Research of the U.S. Department of Energy
under Contract No. DE-AC02-05CH11231.
Support for J.~B.~B.\ was provided by the SciDAC Program of the DOE Office of High Energy Physics
and the Applied Mathematics Program of the DOE Office of
Advanced Scientific Computing Research 
under Contract No. DE-AC02-05CH11231.
At UCSC this research has been supported by the NASA Theory
Program NNX09AK36G, the DOE SciDAC Program (DE-FC02-06ER41438), and
NSF grant AST 0909129.
The computations presented here were performed on the ATLAS Linux
Cluster at LLNL as part of a Grand Challenge Project. 

\bibliographystyle{apj}
\bibliography{ms}

\clearpage

\end{document}